# CERTIFIED SAFE:
## A SCHEMATIC FOR APPROVAL REGULATION OF FRONTIER AI

Cole Salvador
August 2024

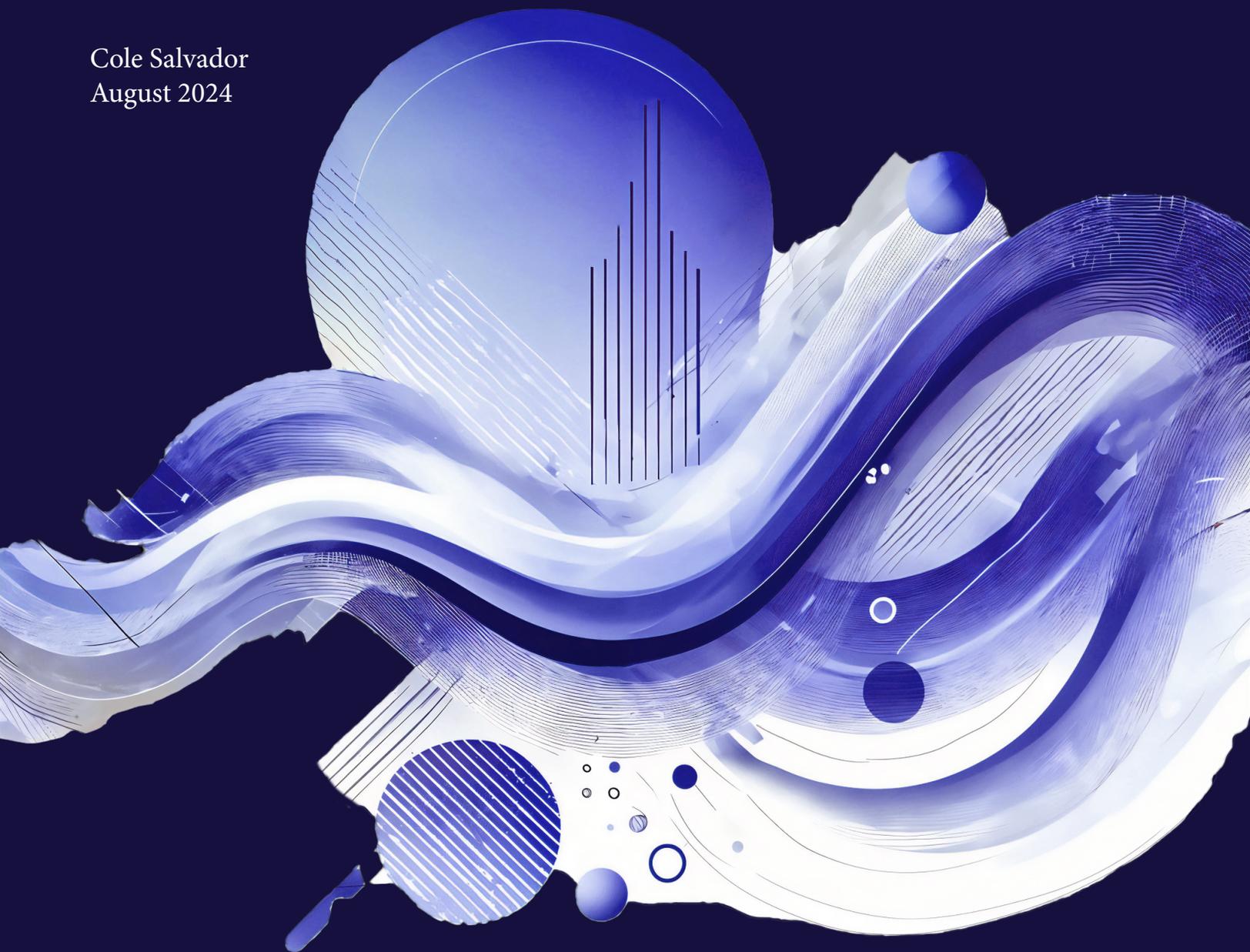

# TABLE OF CONTENTS

# Executive Summary

The frontier of general artificial intelligence (AI) capabilities is progressing rapidly and in potentially dangerous directions. Future AI models could be used to spread misinformation at unprecedented scale, transform warfare, develop bioweapons, or they may become powerful and uncontrollable—a catastrophic combination.

Various structural features amplify these risks of large-scale harm from AI: safety skimping due to intense competition, highly distributed risks despite concentrated rewards, ineffective corporate governance, and massive potential rewards for model theft. This has led to widespread calls for comprehensive, rather than patchwork, regulation of frontier (i.e., the most generally capable) AI.

*Approval regulation* is emerging as a promising regime for such comprehensive regulation. An approval regulation scheme is one in which a firm cannot legally market, or in some cases develop, a product without explicit approval from a regulator on the basis of experiments performed upon the product that demonstrate its safety. For a product to be sold, it must be certified safe. This regime is used to great success by the US Federal Aviation Administration to certify types of planes and the US Food and Drug Administration to certify drugs and medical devices. Many, including AI firm CEOs, researchers, and policymakers, have publicly supported approval regulation of frontier AI.[1]

A number of features of the industry are promising for this regime's prospects. In brief, the current development pipeline for models is standard and maps on well to an approval scheme; the pool of models to regulate is small, given the large investments required to reach dangerous capabilities; approval regulation tackles information asymmetry between firms and the government or public; and national security demands protections against theft of models throughout the development process.

This report proposes a detailed schematic for the approval regulation of frontier AI, primarily based on the FAA's type certification process for aircraft. The FAA process is remarkably successful: American commercial aviation has seen only two fatalities since 2010, despite operating over ten million flights per year.[2]

---

[1] OpenAI CEO Sam Altman: https://www.judiciary.senate.gov/imo/media/doc/2023-05-16_-_qfr_responses_-_altman.pdf; Anthropic CEO Dario Amodei: https://www.judiciary.senate.gov/imo/media/doc/2023-07-26_-_testimony_-_amodei.pdf; Senators Richard Blumenthal and Josh Hawley: https://www.blumenthal.senate.gov/imo/media/doc/09072023bipartisanaiframework.pdf; RAND CEO Jason Matheny: https://www.armed-services.senate.gov/imo/media/doc/23-29_04-19-2023.pdf; Professor Gary Marcus: https://www.judiciary.senate.gov/imo/media/doc/2023-05-16_-_qfr_responses_-_marcus.pdf; and many others.

In this report's proposal, any model planned for training on more computation than some large threshold enters the model certification process. Before training a covered model, the regulator must grant Training Authorization to the applicant firm, which specifies model details, a training timeline, estimates of capability progression, and verification of agreed-upon safety and security measures. During monitored training, a Certification Basis (CB) and certification plan are jointly composed, in which the regulator specifies line item requirements for legal deployment and the applicant plans a set of experiments to meet each one. After training, these experiments are performed, verified, and checked for compliance with requirements. If the CB is satisfied, a model deployment card and instructions for continued safety are issued, specifying legal deployment conditions and required post-deployment safety and security measures.

The implementation of such a process faces five major challenges: (1) noncompliance through unsanctioned deployment (e.g., internally), (2) specification of a demonstrable, specific, and comprehensive list of requirements to guarantee deployment readiness, (3) creation of reliable experiments that demonstrate each requirement, (4) the potential inadequacy of a computation use threshold for filtering out safe models, and (5) minimizing regulatory overhead to preserve American AI innovation.

This report makes three urgent recommendations, which must be heeded earnestly and early to allow for effective approval regulation in the future: agencies, firms, and researchers should improve evaluation techniques; policymakers should begin to specify deployment readiness conditions; and research should bolster compute use tracking and detection.

Four further recommendations, though less urgent, must be followed by the time approval regulation is used if it is to be successful: agencies and firms should research methods to minimize potential regulatory overhead; federal action should establish enhanced whistleblower protections in industry; agencies and researchers should research best model filtering practices to improve on the computational threshold; and the regulator should establish information and personnel security to protect competition and regulator access.

Finally, three general lessons, gleaned from this report's analysis, can apply to any regulatory scheme for frontier AI: regulate not just at deployment, but from early development through post-deployment to protect against theft and noncompliance; consider approval gating in any regime to promote transparency and compliance; and consider the use of checkpoint capability estimation to determine the unpredictability and manageability of training progress.

---

[2] https://www.ntsb.gov/safety/Pages/research.aspx, "2003-2022 Accident Statistics," table 6; https://www.faa.gov/air_traffic/by_the_numbers/media/Air_Traffic_by_the_Numbers_2024.pdf, 6.

# Introduction

In summer 1937, chemist Harold Watkins developed a liquid form of the widely-used strep drug sulfanilamide. After the control lab of his employer, S.E. Massengill Co., cleared the drug on the basis of flavor, appearance, and fragrance tests, 633 crates were sent around the country to be sold, mainly as a prescription treatment for strep throat in children. A month later, following reports from doctors in Oklahoma of deaths, an American Medical Association laboratory became the first to test 'Elixir Sulfanilamide' for toxicity. They found it contained diethylene glycol, a mortally toxic compound used as antifreeze. The US Food and Drug Administration (FDA) dispatched nearly all of its inspectors and chemists to retrieve shipments of the drug. By the time all 234 remaining gallons of the drug were retrieved, the death toll of the sulfanilamide disaster was over 100.[3]

Though Congress had been in favor of legislation on drug safety for some time, the tragedy called lawmakers to action, leading to the 1938 Federal Food, Drug, and Cosmetic Act (FDCA). The FDCA broadened the powers of the FDA, allowing it to recall and seize products, inspect production facilities, and, most importantly, require firms to submit evidence of new drug safety in order to gain approval to sell to the public.[4] Yet even the expanded powers of the FDCA did not satisfy public safety concerns. It allowed the FDA only 60 days to deny unsafe drugs and prevented regulators from performing or verifying efficacy tests on new drugs.[5] It took another tragedy—this time, primarily abroad[6]—for the Kefauver-Harris amendments of 1962 to empower the FDA with 180 days-to-deny and new drug clinical trials for safety and effectiveness. In the decades since, the FDA's powers have expanded even further. Today, FDA-regulated products account for 20 percent of all US consumer spending.[7] It has been called the "world's most powerful regulatory agency."[8]

The US may today be in a moment not unlike the quiet before Watkins's development of Elixir Sulfanilamide. This time, however, the looming threat is posed by frontier artificial intelligence (AI). Like drug development in the 1930s, regulation of AI is largely left to development firms themselves. Any pre-market tests performed by a developer are voluntary. And though there is plenty of appetite in Congress and elsewhere for comprehensive regulation of frontier AI,[9] the only current measures in the US are reporting requirements, set out in an executive order, that do not apply to current models.[10] In many labs around the country today, scientists—this time of machine learning, rather than chemistry—are working to develop increasingly capable AI. Some, however, have stopped work to warn of dangers associated with that development.[11] With experts warning of risk and minimal external oversight, no one can confidently say that the status quo will safely navigate all threats on the path towards the promise of valuable AI.

---

[3] Carol Ballentine, "Sulfanilamide Disaster," *FDA Consumer*, June 1981, https://www.fda.gov/files/about%20fda/published/The-Sulfanilamide-Disaster.pdf.
[4] Clinton Lam and Preeti Patel, "Food, Drug, and Cosmetic Act," StatPearls, July 31, 2023, https://www.ncbi.nlm.nih.gov/books/NBK585046/.
[5] Julian West, "The Accidental Poison that Founded the Modern FDA," *The Atlantic*, January 16, 2018, https://www.theatlantic.com/technology/archive/2018/01/the-accidental-poison-that-founded-the-modern-fda/550574/.
[6] Though most of the effects of the thalidomide disaster were abroad, 17 abnormal births in the US were caused by thalidomide. "Kefauver-Harris Amendments Revolutionized Drug Development," US Food and Drug Administration, October 2012, https://www.gvsu.edu/cms4/asset/F51281F0-00AF-E25A-5BF632E-8D4A243C7/kefauver-harris_amendments.fda.thalidomide.pdf.
[7] Merlin Stein and Connor Dunlop, *Safe Before Sale: Learnings From the FDA's Model of Life Sciences Oversight for Foundation Models*, Emerging Technology & Industry Practice (United Kingdom: Ada Lovelace Institute, 2023), 5, https://www.adalovelaceinstitute.org/wp-content/uploads/2023/12/2023_12_ALI_Safe-before-sale_Discussion_paper.pdf.
[8] Daniel Carpenter, *Reputation and Power: Organizational Image and Pharmaceutical Regulation at the FDA* (Princeton: Princeton University Press, 2010), 22.

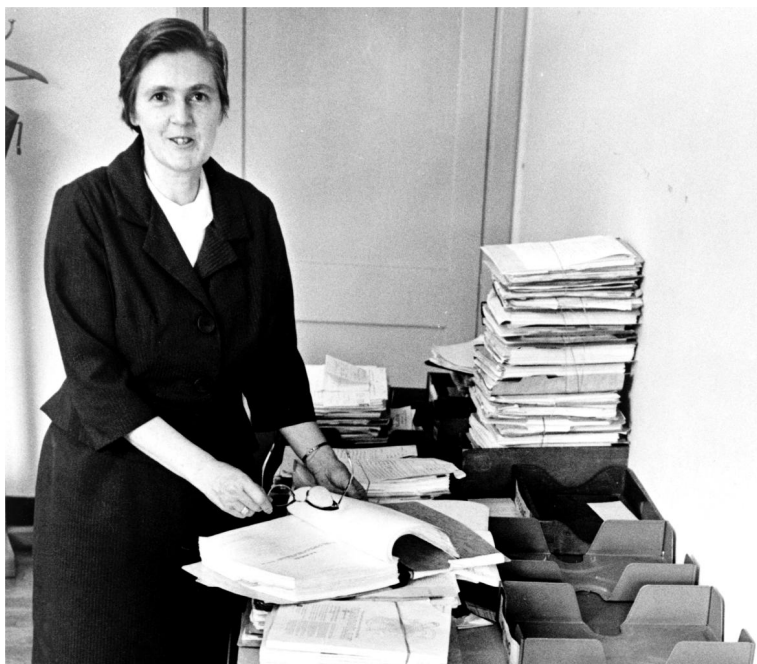

*Frances Kelsey, pictured here in the 1960s, is the medical officer who both discovered the toxic compound in Elixir Sulfanilamide in 1937 and, twenty-five years later, rejected the drug application for thalidomide, which was later found to cause birth defects. (Wikimedia Commons)*

Today, any firm in the US can train any AI model at any time, of any size, with any number of computers, accepting any level of risk,[12] under any circumstances.[13] The current development practices of major AI firms—often inspired by the Silicon Valley mantra "Move fast and break things"[14]—are startlingly dissimilar to the rigorous safety requirements characteristic of high-risk industries, which frontier AI has recently become. What would it look like if, instead, developers were required to demonstrate the safety of their most capable models before deploying them widely to the public? This report considers one such proposal: pre-development and pre-deployment approval regulation for frontier AI. Through a schematic for such a proposal, this report shows that scrutiny of safety and security of the highest-risk models maps well onto the current development pipeline for frontier AI. Next, it presents the major challenges to the implementation of such a policy, observing both over- and under-explored obstacles of varying levels of solubility, including approval regulation in the context of alternative regimes for regulating AI. Finally, the report details recommendations for relevant stakeholders and outlines the most promising areas for future work.

---

[9] See Richard Blumenthal and Josh Hawley, "Bipartisan Framework for U.S. AI Act," September 7, 2023, https://www.blumenthal.senate.gov/imo/media/doc/09072023bipartisanaiframework.pdf; "Majority Leader Schumer Delivers Remarks To Launch SAFE Innovation Framework For Artificial Intelligence at CSIS," Senate Democrats, June 21, 2023, https://www.democrats.senate.gov/news/press-releases/majority-leader-schumer-delivers-remarks-to-launch-safe-innovation-framework-for-artificial-intelligence-at-csis; Mitt Romney, Jack Reed, Jerry Moran, and Angus King, "Letter to Senators Schumer, Rounds, Heinrich, and Young," April 16, 2024, https://www.romney.senate.gov/wp-content/uploads/2024/04/240415-AI-Letter-final.pdf; "SB 1047: Safe and Secure Innovation for Frontier Artificial Intelligence Models Act, Amended Assembly (7/3/2024)" Digital Democracy - CalMatters, July 3, 2024, https://digitaldemocracy.calmatters.org/bills/ca_202320240sb1047; Tharin Pillay and Harry Booth, "Exclusive: Renowned Experts Pen Support for California's Landmark AI Safety Bill," *Time*, August 7, 2024, https://time.com/7008947/california-ai-bill-letter/.

[10] These measures "do not apply to current models" because no model trained as of writing passes the compute reporting threshold. Exec. Order No. 14110, 88 Fed. Reg. 75191 (2023), https://www.federalregister.gov/documents/2023/11/01/2023-24283/safe-secure-and-trustworthy-development-and-use-of-artificial-intelligence.

[11] See Cade Metz, "'The Godfather of AI' Quits Google and Warns of Danger Ahead," *New York Times*, May 1, 2024, https://www.nytimes.com/2023/05/01/technology/ai-google-chatbot-engineer-quits-hinton.html; "Statement on AI Risk," Center for AI Safety, accessed July 24, 2024, https://www.safe.ai/work/statement-on-ai-risk; Melissa Heikkilä, "AI is at an Inflection Point, Fei-Fei Li Says," *MIT Technology Review*, November 14, 2023, https://www.technologyreview.com/2023/11/14/1083352/ai-is-at-an-inflection-point-fei-fei-li-says/.

[12] Although firms are subject to some negligence liability, this can be thought of as an additional component of risk, amplifying costs to the firm.

[13] Legally, that is. The capital expenditures required to train frontier AI are large, on the order of a hundred million dollars (and growing), which is prohibitive for most firms.

[14] See Jonathan Taplin, *Move Fast and Break Things: How Facebook, Google, and Amazon Cornered Culture and Undermined Democracy* (Little, Brown, 2017).

## Why Regulate Frontier AI?

The frontier of general AI capabilities is expanding rapidly. (Throughout this report, any AI model which is generally capable and near the state-of-the-art in terms of performance qualifies as "frontier AI."[15]) Since the introduction of the transformer architecture—the prevailing language modeling algorithm behind all current frontier AI—in 2017,[16] total computation use in model training has grown over four times per year.[17] Algorithmic progress—new machine learning algorithms and engineering improvements that improve capabilities—has proceeded similarly quickly. In addition, the investment into frontier AI training has doubled every nine months. By 2017, the largest training investment was $600K; today, the most capable models cost over a hundred million dollars to train. If trends hold, one billion dollars will be expended to train a single model by 2027.[18]

These incredible investments have not been without capability rewards. No model trained before 2020 achieves performance significantly better than random guessing on MMLU, a multiple choice dataset that is the most widely-used benchmark for frontier AI performance. Today, the most capable models exceed the performance of even the most capable human experts.[19] But rapid progress also surfaces hints of the risks associated with the technology. For instance, various groundbreaking capabilities in general AI appear to 'emerge' from thin air. These so-called "emer-gent capabilities" are unpredictable 'jumps' in model ability, for example to solve difficult mathematics problems.[20] Emergent capabilities present a significant risk for model safety, given unpredictable and rapidly-appearing behavior can be difficult to protect against.[21] This is only one among the many technical safety issues presented by unfettered frontier AI development.

---

[15] In this context, "generally capable" refers to foundation models. A "foundation model is any model that is trained on broad data that can be adapted to a wide range of downstream tasks," following Rishi Bommasani et al., "On the Opportunities and Risks of Foundation Models," arXiv, 2022, 3, https://doi.org/10.48550/arXiv.2108.07258. Current models which qualify as "frontier AI" include GPT-4 (OpenAI, "GPT-4 Technical Report," arXiv, 2024, https://doi.org/10.48550/arXiv.2303.08774); Claude 3 (Anthropic, "The Claude 3 Model Family: Opus, Sonnet, Haiku," 2024, https://www-cdn.anthropic.com/de8ba9b01c9ab7cbabf-5c33b80b7bbc618857627/Model_Card_Claude_3.pdf); and Gemini 1 (Gemini Team, Google, "Gemini: A Family of Highly Capable Multimodal Models," arXiv, 2024, https://doi.org/10.48550/arXiv.2312.11805).
[16] Introduced in Ashish Vaswani et al., "Attention is all you need," *Advances in Neural Information Processing Systems* 30 (2017).
[17] "Machine Learning Trends: Investment Trends," Epoch AI, June 7, 2024, https://epochai.org/trends#investment-trends-section.
[18] AlphaGo Zero cost $600K to train, while Gemini 1.0 Ultra and GPT-4 cost around $100M. Trends and prediction from Ben Cottier et al., "The Rising Costs of Training Frontier AI Models," arXiv, 2024, 6, https://doi.org/10.48550/arXiv.2405.21015.
[19] GPT-2-XL, trained before 2020, achieved 32.4% accuracy on MMLU (random guessing yields 25% accuracy). Today, Gemini Ultra achieves 90% accuracy. "Multi-task Language Understanding on MMLU," Papers With Code, accessed July 24, 2024, https://paperswithcode.com/sota/multi-task-language-understanding-on-mmlu. Unspecialized human performance is estimated at 34.5%, while an operationalization of "expert performance" that entails assigning a domain expert to each question according to that question's topic achieves 89.8%. Dan Hendrycks et al., "Measuring Massive Multitask Language Understanding," arXiv, 2021, 3, https://doi.org/10.48550/arXiv.2009.03300.

There exist significant present harms caused by the technology that will only increase in scale and intensity as models grow more capable. Current models, for example, have been shown to exhibit "covert stereotypes that are more negative than any human stereotypes about African Americans ever experimentally recorded."[22] Image generation abilities not only often give racist and sexist results,[23] but can also be used to create harmful "Deepfakes": images (or audio and video) of real people.[24] Experts warn of Deepfakes being used to create pornographic materials without the knowledge of those depicted[25] or undermine the democratic process through widespread misinformation, such as fabricated political speeches.[26] Current models are also being used as effective cybercriminals to enhance, economize, and scale up phishing and other scams.[27] In addition, all of these risks may disproportionately affect vulnerable communities.[28]

Models already frequently fail. In stunning, unexpected, and undesirable ways, they demonstrate an inability to be controlled.[29] And yet, the dangers surfaced thus far may be only the tip of the iceberg. What happens when models are capable enough that their failures cause unacceptable, large-scale harms?

---

[20] See Jason Wei et al., "Emergent Abilities of Large Language Models," arXiv, 2022, https://doi.org/10.48550/arXiv.2206.07682; Zhengxiao Du et al., "Understanding Emergent Abilities of Language Models from the Loss Perspective," arXiv, 2024, https://doi.org/10.48550/arXiv.2403.15796. There exists some discussion as to whether these discontinuities in capability emergence are *theoretically* valid, but due to the frequency of evaluation and the lack of perfect evaluation suites, the problem is certainly pragmatically relevant. For criticism, see Rylan Schaeffer, Brando Miranda, and Sanmi Koyejo, "Are Emergent Abilities of Large Language Models a Mirage?" arXiv, 2023, https://doi.org/10.48550/arXiv.2304.15004. Some recent work calls this criticism into question. See Core Francisco Park et al., "Emergence of Hidden Capabilities: Exploring Learning Dynamics in Concept Space," arXiv, 2024, https://doi.org/10.48550/arXiv.2406.19370.
[21] See discussion of "The Unexpected Capabilities Problem" in Markus Anderljung et al., "Frontier AI Regulation: Managing Emerging Risks to Public Safety," arXiv, 2023, §2.2.1, https://doi.org/10.48550/arXiv.2307.03718.
[22] Valentin Hofmann et al., "Dialect Prejudice Predicts AI Decisions About People's Character, Employability, and Criminality," arXiv, 2024, 1, https://doi.org/10.48550/arXiv.2403.00742.
[23] See Ananya, "AI Image Generators Often Give Racist and Sexist Results: Can They Be Fixed?" *Nature* 627 (2024): 722-5, https://doi.org/10.1038/d41586-024-00674-9.
[24] See Thanh Thi Nguyen et al., "Deep Learning for Deepfakes Creation and Detection: A Survey," *Computer Vision and Image Understanding* 223 (2022), https://doi.org/10.1016/j.cviu.2022.103525.
[25] See, e.g., Andjela Milivojevic, "'Undressed' by AI: Serbian Women Defenceless Against Deepfake Porn," *Balkan Insight*, July 3, 2024, https://balkaninsight.com/2024/07/03/undressed-by-ai-serbian-women-defenceless-against-deepfake-porn/.
[26] Nilay Patel, "Harvard Professor Lawrence Lessig on Why AI and Social Media Are Causing a Free Speech Crisis for the Internet," *The Verge*, October 24, 2023, https://www.theverge.com/23929233/lawrence-lessig-free-speech-first-amendment-ai-content-moderation-decoder-interview.
[27] See Fredrik Heiding et al., "Devising and Detecting Phishing: Large Language Models vs. Smaller Human Models," arXiv, 2023, https://doi.org/10.48550/arXiv.2308.12287; Daniel Kelley, "WormGPT - The Generative AI Tool Cybercriminals Are Using to Launch Business Email Compromise Attacks," *Slash Next* (blog), July 13, 2023, https://slashnext.com/blog/wormgpt-the-generative-ai-tool-cybercriminals-are-using-to-launch-business-email-compromise-attacks/.
[28] See Patricia Costinhas, "AI's Adverse Effects on Marginalized Communities," *Impakter*, August 23, 2023, https://impakter.com/ai-adverse-effects-on-marginalized-communities/.
[29] See Kevin Roose, "A Conversation With Bing's Chatbot Left Me Deeply Unsettled," *New York Times*, February 17, 2023, https://www.nytimes.com/2023/02/16/technology/bing-chatbot-microsoft-chatgpt.html; Nick McKenna et al., "Sources of Hallucination by Large Language Models on Inference Tasks," arXiv, 2023, https://doi.org/10.48550/arXiv.2305.14552; Maksym Andriushchenko, "Jailbreaking Leading Safety-Aligned LLMs with Simple Adaptive Attacks," arXiv, 2024, https://doi.org/10.48550/arXiv.2404.02151.

Experts and development firms alike warn of such future dangers associated with frontier AI if left unregulated. Perhaps the most striking example of this is the Center for AI Safety's open letter, which states in full: "Mitigating the risk of extinction from AI should be a global priority alongside other societal-scale risks such as pandemics and nuclear war."[30] The signatory list is a who's-who of AI: two founders of the field, the CEOs of the three leading American AI firms, policymakers, professors, and hundreds of other experts.[31] In addition, a 2024 article in *Science*, from many of the field's leading experts, warns of "risks that include large-scale social harms, malicious uses, and an irreversible loss of human control over autonomous AI systems."[32] A 2023 survey of researchers who publish at prestigious AI conferences yielded similarly startling results.[33] The most involved actors are extremely worried about the prospect of large-scale future risks from frontier AI.

Further, various features of the broader development landscape of frontier AI present challenges for safe development. First, competitive pressures between domestic developers reward skimping on safety measures and pushing out products before they have been adequately vetted.[34] The primary cause of pernicious competition in this case is the massive projected reward for reaching economically-valuable general AI before one's competitors. This issue is exacerbated by an increasing number of relevant developers. These facts, taken together, indicate that an unmitigated race towards valuable AI may not benefit the public interest.

Second, potential large-scale risks from general AI are highly distributed across large populations, meaning users and developers do not bear the brunt of consequence for failures, despite reaping the lion's share of rewards for success. This market failure appears to be addressable primarily through direct government regulation.[35]

Third, questions have been raised about the corporate structures of leading developers. Despite some developers ostensibly operating as nonprofit or qualified for-profit (e.g., public benefit) corporations, billions of dollars of investment from for-profit firms including Microsoft, Google, and Amazon have poured in.[36] Indeed, OpenAI, which is governed by a nonprofit board, has recently considered "chang-ing its governance structure to a for-profit business that the firm's nonprofit board doesn't control."[37] Given the risks—and massive projected rewards—associated with frontier AI development, an eye towards profit may entrench competitive pressures. In a recent instance, OpenAI "pressured" its safety team to "speed through" model safety testing in order to meet a launch date—precisely the sort of skimping on safety that competition incentivizes.[38] All of these factors increase the likelihood that transformative AI is developed irresponsibly, without great care for societal effects, and with massive risks.

---

[30] Center for AI Safety, "Statement on AI Risk."
[31] In order, the "founders" (of deep learning) and "CEOs" are Geoffrey Hinton (University of Toronto), Yoshua Bengio (University of Montreal), Demis Hassabis (Google DeepMind), Sam Altman (OpenAI), and Dario Amodei (Anthropic). The policymakers include Congressman Ted Lieu, former Taiwan Minister of Digital Affairs Audrey Tang, and others. Other notable figures include Bill Gates, Harvard Law Professors Laurence Tribe and Lawrence Lessig, Jaan Tallinn, and many others.
[32] Yoshua Bengio et al., "Managing Extreme AI Risks Amid Rapid Progress," *Science* 384, no. 6689 (May 2024): 842, https://doi.org/10.1126/science.adn0117.
[33] See Katja Grace et al., "Thousands of AI Authors on the Future of AI," arXiv, 2024, https://doi.org/10.48550/arXiv.2401.02843.
[34] Stuart Armstrong, Nick Bostrom, and Carl Shulman, "Racing to the Precipice: a Model of Artificial Intelligence Development," October, 2013, https://www.fhi.ox.ac.uk/wp-content/uploads/Racing-to-the-precipice-a-model-of-artificial-intelligence-development.pdf.
[35] Alternative theories are discussed further in the Implementation Challenges section.
[36] See "Amazon and Anthropic Deepen Their Shared Commitment to Advancing Generative AI," Amazon, March 27, 2024, https://www.aboutamazon.com/news/company-news/amazon-anthropic-ai-investment; "Google Invest In Anthropic For $2 Billion As AI Race Heats Up," Forbes, October 31, 2023, https://www.forbes.com/sites/qai/2023/10/31/google-invests-in-anthropic-for-2-billion-as-ai-race-heats-up/; Jordan Novet, "Microsoft's $13 Billion Bet on OpenAI Carries Huge Potential Along with Plenty of Uncertainty," April 9, 2023, https://www.cnbc.com/2023/04/08/microsofts-complex-bet-on-openai-brings-potential-and-uncertainty.html.

Unregulated AI development also poses a threat to US national security. It is already abundantly clear from the focus of US chip export controls that foreign AI progress is a cause for national security concern. The stringent controls of 2022, tightened in 2023, are designed explicitly to prevent US adversaries from advancing AI capabilities.[39] "Advanced AI capabilities," the 2023 control press release states, "present U.S. national security concerns because they can be used to improve the speed and accuracy of military decision making, planning, and logistics."[40] Generally-capable frontier AI is clearly implicated in these concerns: military planning and advising tools exist which are built around the general capabilities of frontier models.[41] Increasingly capable models are likely to play a more substantial role in the future of warfare.[42] In addition, work from OpenAI, RAND, and Anthropic indicates that next generation models may lower the barrier to entry for the creation of dangerous pathogens.[43] This risks more sophisticated or frequent attacks from state and non-state actors alike.[44]

Were the models trained by leading firms available only to American organizations, these may be less troublesome developments. However, the frontier models trained and deployed today are not subject to security or limited deployment requirements. The lack of deployment requirements means that almost all models are widely available for use. Although use purchased through an AI firm is typically subject to some monitoring and safety guardrails, there exist no standardized requirements or practices. Indeed, "jailbreaking" models to elicit dangerous capabilities despite safety measures is hardly a challenge; there are many new attacks every month (even in the public academic literature) that are still effective.[45]

---

[37] "OpenAI CEO Says Company Could Become For-Profit Corporation, The Information Reports," *Reuters*, June 15, 2024, https://www.reuters.com/technology/artificial-intelligence/openai-ceo-says-company-could-become-benefit-corporation-information-2024-06-15/.
[38] Pranshu Verma, Nitasha Tiku, and Cat Zakrzewski, "OpenAI Promised to Make Its AI Safe. Employees Say It 'Failed' Its First Test," *Washington Post*, July 12, 2024, https://www.washingtonpost.com/technology/2024/07/12/openai-ai-safety-regulation-gpt4/.
[39] See Gregory C. Allen, "Choking Off China's Access to the Future of AI," October 11, 2022, https://www.csis.org/analysis/choking-chinas-access-future-ai.
[40] "Commerce Strengthens Restrictions on Advanced Computing Semiconductors, Semiconductor Manufacturing Equipment, and Supercomputing Items to Countries of Concern," Bureau of Industry and Security Office of Congressional and Public Affairs, October 17, 2023, https://www.bis.doc.gov/index.php/documents/about-bis/newsroom/press-releases/3355-2023-10-17-bis-press-release-acs-and-sme-rules-final-js/file.
[41] For example, Scale AI offers "Donovan," a "Digital Staff Officer for national security," which is powered by frontier language models. "Donovan: AI Digital Staff Officer for National Security," Scale AI, accessed July 24, 2024, https://scale.com/donovan. This product is similar to Cognition AI's "Devin," which creates similar scaffolding around language models but is applied for software engineering rather than military advising. Scott Wu, "Introducing Devin, the First AI Software Engineer," *Cognition AI blog*, March 12, 2024, https://www.cognition.ai/blog/introducing-devin.
[42] See Mary L. Cummings, "Artificial Intelligence and the Future of Warfare," ISBN 9781784131982 (London, UK: Chatham House, 2017), https://www.chathamhouse.org/sites/default/files/publications/research/2017-01-26-artificial-intelligence-future-warfare-cummings-final.pdf.
[43] See Christopher A. Mouton, Caleb Lucas, and Ella Guest, *The Operational Risks of AI in Large-Scale Biological Attacks: Results of a Red-Team Study*, RR-A2977-1 (Santa Monica, CA: RAND, 2023), https://www.rand.org/pubs/research_reports/RRA2977-2.html; "Frontier Threats Red Teaming for AI Safety," Anthropic, July 26, 2023, https://www.anthropic.com/news/frontier-threats-red-teaming-for-ai-safety; Tejal Patwardhan et al., "Building an Early Warning System for LLM-Aided Biological Threat Creation," January 31, 2024, https://openai.com/index/building-an-early-warning-system-for-llm-aided-biological-threat-creation/#_4SUgyekFkwzOK2Vq8ES0pq.
[44] See "Fact Sheet: DHS Advances Efforts to Reduce the Risks at the Intersection of Artificial Intelligence and Chemical, Biological, Radiological, and Nuclear (CBRN) Threats," Department of Homeland Security, accessed July 24, 2024, https://www.dhs.gov/sites/default/files/2024-04-24_0429_cwmd-dhs-fact-sheet-ai-cbrn.pdf.

The largest risks, however, are presented when a well-resourced actor, such as a foreign state, gains full access to a highly capable model. In general, once a model is trained, its so-called 'weights' are stored in what amounts to a single computer file somewhere in the development lab.[46] Gaining access to a model's weights is functionally equivalent to having trained the model oneself—without having to spend hundreds of millions of dollars. Thus, the most worrying possibility is that a frontier AI firm in the US trains a highly capable model and it is stolen by a well-resourced actor. Unfortunately, recent research by RAND on "Securing AI Model Weights" indicates that current security practices are insufficient and even large investments in security may not be able to defend against attempts by the most resourced actors (e.g., states) to steal models.[47] Even merely trained models (i.e., ones not yet in use) can be stolen in this way. Given the US's current lead on AI, theft of a frontier model, perhaps of a future generation even more applicable to a military context, would be disastrous. This report shows that before-training approval[48] can be used to protect against such theft.

Clearly, then, the risks of frontier AI are unlikely to be adequately mitigated without comprehensive government regulation. As this report will outline, a regulatory scheme that allows for development or deployment only when safety has been reasonably demonstrated has the potential to reduce risk along each of these dimensions. Unexpected capability improvements, present and future harms, pernicious pressures between developers, and security risks from state and non-state actors wielding AI can all be protected against under such a scheme. The backbone of this protection is the regime of approval regulation.

## What is Approval Regulation?

For the purposes of this report, approval regulation is a regulation scheme in which a firm cannot legally market, or in some cases develop, a product without explicit approval from a regulator on the basis of experiments performed upon the product that demonstrate its safety.[49] In other words, for a product to be sold, it must be certified safe. This scheme is the one used by the US Federal Aviation Administration (FAA) and Food and Drug Administration (FDA), two of the most successful regulatory agencies in the world.[50]

---

[45] See Raz Lapid, Ron Langberg, and Moshe Sipper, "Open Sesame! Universal Black Box Jailbreaking of Large Language Models," arXiv, 2023, https://doi.org/10.48550/arXiv.2309.01446; Xuandong Zhao et al., "Weak-to-Strong Jailbreaking on Large Language Models," arXiv, 2024, https://doi.org/10.48550/arXiv.2401.17256. One attack even easily bypasses the most strictly controlled form of deployment access, a firm's own API: Danny Halawi et al., "Covert Malicious Finetuning: Challenges in Safeguarding LLM Adaptation," arXiv, 2024, https://doi.org/10.48550/arXiv.2406.20053.
[46] A lab may also retain multiple copies of the file. A model's "weights" are a massive set of numbers which determine the its behavior.
[47] Sella Nevo et al., *Securing AI Model Weights: Preventing Theft and Misuse of Frontier Models*, RR-A2849-1 (Santa Monica, CA: RAND, 2024), https://www.rand.org/pubs/research_reports/RRA2849-1.html.
[48] This report uses the less elegant "before-training" rather than "pre-training" because pre-training (or simply "pretraining") is used in machine learning to refer to a specific phase of model training (which is not, in actuality, *prior to* training).

The FAA's primary approval regulation process, known as the type certification process, is used to approve new models of airplanes for production and flight.[51] The process begins when an aviation firm (such as Boeing or Airbus) consults with the FAA to propose a plane design. After a formal application is submitted, the regulator and applicant design and implement tests on prototype aircraft which, taken together, can be used to demonstrate the airworthiness and overall safety of the plane in line with federal regulations. These tests include ground tests, such as load-bearing and extreme conditions testing, and flight tests in a wide range of conditions.[52] Once the FAA verifies that compliance has been adequately demonstrated, approval is awarded to the firm to produce and sell aircraft of the specified design.[53]

The FDA's approval regulation scheme is used to certify pharmaceuticals for public sale.[54] After discovering promising compounds, preclinical studies on dosing and toxicity are performed under the supervision of FDA inspectors and experts. Multiple phases of large-scale clinical trials (involving thousands of patients) are overseen by the firm and FDA throughout design, implementation, and results. A new drug application is then submitted by the firm, which includes all experimental results and drug information subject to rigorous documentation guidelines. A review team of FDA experts then has 6 to 10 months to decide whether to approve the drug, during which time inspectors and experts have wide latitude to investigate reported information and arguments of safety and efficacy.[55] Both the FDA and FAA processes expend significant effort on post-approval monitoring to ensure safety through a product's lifespan.

---

[49] Note that in some industries the mentioned empirical case will be required to demonstrate more than safety. For example, the FDA process requires drugs to also demonstrate their efficacy. See Daniel Carpenter, and Michael M. Ting, "Regulatory Errors with Endogenous Agendas," *American Journal of Political Science* 51, no. 4 (October 2007): 835-52; Daniel Carpenter, Justin Grimmer, and Eric Lomazoff, "Approval Regulation and Endogenous Consumer Confidence: Theory and Analogies to Licensing, Safety, and Financial Regulation," *Regulation & Governance* 4, no. 4 (December 2010): 383-407, https://doi.org/10.1111/j.1748-5991.2010.01091.x; Daniel Carpenter, "Protection Without Capture: Product Approval by a Politically Responsive, Learning Regulator," *American Political Science Review* 98, no. 4 (November 2004): 613-31, doi:10.1017/S0003055404041383.

[50] See "A Brief History of the FAA," Federal Aviation Administration, accessed July 24, 2024, https://www.faa.gov/about/history/brief_history; Thomas R. Fleming, David L. Demets, and Lisa M. McShane, "Discussion: the Role, Position, and Function of the FDA—the Past, Present, and Future," *Biostatistics* 18, no. 3 (2017): 417-21, https://doi.org/10.1093/biostatistics/kxx023.

[51] The process is also used to certify various other aviation units, such as helicopters, blimps, engines, propellers, etc.

[52] "Test and Certification: Approvals for Entry-in-Service," Airbus, accessed July 24, 2024, https://www.airbus.com/en/products-services/commercial-aircraft/the-life-cycle-of-an-aircraft/test-and-certification. There are many other types of tests also used to find compliance, such as tests on individual plane components.

[53] See Federal Aviation Administration, *Type Certification*, FAA Order 8110.4C Chg 7 (Washington, DC: Federal Aviation Administration, 2023), https://www.faa.gov/documentLibrary/media/Order/Order_8110.4C_CHG_7.pdf.

[54] A slightly modified process is also used to certify medical devices and software. This section focuses on pharmaceuticals for simplicity.

[55] "Step 4: FDA Drug Review," US Food and Drug Administration, January 4, 2018, https://www.fda.gov/patients/drug-development-process/step-4-fda-drug-review.

Approval regulation is on the table for frontier AI development, which as previously discussed is now a high-risk industry. Indeed, approval regulation for AI has been proposed by many relevant stakeholders (in addition to the congressional proposals mentioned previously).[56] OpenAI CEO Sam Altman testified before the US Senate that his company supports "pre-deployment risk assessments" for "licensing requirements… common in safety-critical and other high-risk contexts."[57] RAND CEO Jason Matheny similarly testified before the US Senate in favor of licensing: "we need a licensing regime… and we need to have certain guarantees of security before [models] are deployed."[58] The "Bipartisan Framework for U.S. AI Act," proposed by Senators Blumenthal and Hawley, calls for licensing requirements that are "conditioned on developers maintaining risk management, pre-deployment testing, data governance, and adverse incident reporting programs" and provide to regulators "the authority to conduct audits of companies seeking licenses."[59] Various work has floated views on the prospects of approval regulation for AI,[60] but no work has comprehensively outlined a detailed schematic while assessing the feasibility and effectiveness of such a proposal.[61] This report aims to do just that.

## The Prospects of Approval Regulation for Frontier AI

Various facts of frontier AI development make approval regulation a particularly promising regime to facilitate safe, trustworthy, and swift innovation. For one, the current frontier AI development pipeline is standardized across developers and maps particularly well onto the regulatory checkpoints of approval regulation. This means that developers will not be required to significantly alter any part of their development processes to comply with regulators. At an appropriate level of abstraction, frontier AI model development is like new airplane development: projects begin with conceptual and engineering design, a few months of production and steep capital expenditures yield a testable prototype, experiments performed on the prototype indicate its usefulness and safety, there is a deployment phase when the product becomes available to the public, and continued monitoring of the product is performed over its lifecycle to ensure reliability and safety. These well-defined phases to development have allowed regulators to construct the 'gates' that make FAA ap-proval regulation effective. The same principles can be applied in the case of frontier AI.

---

[56] Not all of the congressional proposals are strictly approval regulation schemes per se.
[57] *Oversight of A.I.: Rules for Artificial Intelligence*, 118th Cong. (2023) (statement of Samuel Altman, OpenAI CEO), https://www.judiciary.senate.gov/imo/media/doc/2023-05-16_-_qfr_responses_-_altman.pdf.
[58] *Hearing To Receive Testimony on the State of Artificial Intelligence and Machine Learning Applications to Improve Department of Defense Operations*, 118th Cong. (2023) (statement of Jason Matheny, RAND President and CEO), https://www.armed-services.senate.gov/imo/media/doc/23-29_04-19-2023.pdf. See also Olaf J. Groth, Mark J. Nitzberg, and Stuart J. Russell, "AI Algorithms Need FDA-Style Drug Trials," *Wired*, August 15, 2019, https://www.wired.com/story/ai-algorithms-need-drug-trials/.
[59] Blumenthal and Hawley, "Bipartisan Framework."
[60] See Andrew Tutt, "An FDA for Algorithms," *Administrative Law Review* 69, no. 1 (2017): 83-123, https://dx.doi.org/10.2139/ssrn.2747994; Daniel Carpenter and Carson Ezell, "An FDA for AI? Pitfalls and Plausibility of Approval Regulation for Frontier Artificial Intelligence," arXiv, 2024, https://doi.org/10.48550/arXiv.2408.00821; Anna Lenhart and Sarah Myers West, *Lessons from the FDA for AI*, AI Now Institute, August 1, 2024, https://ainowinstitute.org/wp-content/uploads/2024/08/20240801-AI-Now-FDA.pdf.
[61] The most intensive existing treatment of this issue is Stein and Dunlop, *Safe Before Sale*. This work focuses on lessons from the FDA approval process. It does not assess the prospects of approval regulation for AI or present a detailed model, though it generates many useful recommendations for those seeking to implement effective approval regulation. A more authority-focused account is given in Lenhart and Myers West, *Lessons from the FDA for AI*. For a more academic discussion, see Carpenter and Ezell, "An FDA for AI?"

Second, approval regulation is particularly potent in industries with extremely low risk tolerance. This is because regulators have fine control over which products can be deployed widely and under which circumstances. Since this proposal seeks to mitigate the most severe large-scale risks, tolerance is necessarily low.

Third, only the most highly capable models pose significant large-scale risks. A typical significant challenge to approval regulation is the talent and funding needed to adequately scrutinize each application reasonably quickly. Fortunately, the pool of models that regulators see will be small, since simple computational thresholds[62] (such as the $10^{26}$ floating point operations threshold used in the US Executive Order on AI[63]) can filter out all but the most capable models before incurring regulatory expenditures. This means that regulatory scrutiny will be more effective, swift, and economical.

Fourth, approval regulation is designed to combat information asymmetry, which is characteristic of current AI development. Since this is a largely unregulated industry, frontier AI developers are not particularly forthcoming with information relevant to model safety. While protecting proprietary information,[64] approval regulators encourage developers to share safety information in order to become trusted and make the case for certification of their products. This benefits the public and economy by generating consumer knowledge.[65] It also shifts developer incentives towards creating safe products and effective methods for generating safety-substantiating data.

Finally, the significant risks associated with widespread deployment of frontier AI, including post-deployment model changes, seem to require a regulator with the powers of approval. Because new behaviors can surface over a model's deployment lifetime,[66] regulators should be empowered to order the recall of a product if necessary. For example, in March 2024 the FAA grounded all Boeing 737-9 MAX aircraft and halted their production until inspections could be performed.[67] Recall orders may be particularly useful in frontier AI, where developers display different risk tolerance than regulators and may benefit from continuing to deploy unsafe models.[68] Any regulator which wields the power to force product recalls would also have de facto control over pre-market deployment decisions, meaning it is an approval regulator in all but name. Any approval regulator of this type is also unlikely to be effective on a state-by-state basis, meaning legislation at the federal level is required.

---

[62] See Lennart Heim, "Training Compute Thresholds: Features and Functions in AI Governance," arXiv (working paper), 2024, https://doi.org/10.48550/arXiv.2405.10799.
[63] Exec. Order No. 14110.
[64] Information protection is extremely important for a successful approval regulation scheme. The FAA, for example, has explicit provisions in its Freedom of Information Act Program which rigorously protect proprietary information, including "description, design, and substantiating data received from applicants." Federal Aviation Administration, *Type Certification*, 48. This is especially true of frontier AI, where institutional knowledge and intellectual property form a significant portion of a firm's value. The FAA has been successful at this information security task: though aircraft designs are expensive (production costs can run into the hundreds of millions as well), they are not regularly stolen.
[65] Carpenter, Grimmer, and Lomazoff, "Approval Regulation and Endogenous Consumer Confidence," 383.
[66] For example through federated learning (Felix Wagner et al., "Post-Deployment Adaptation with Access to Source Data via Federated Learning and Source-Target Remote Gradient Alignment," arXiv, 2023, https://doi.org/10.48550/arXiv.2308.16735) or online adaptation (Amal Rannen-Triki et al., "Revisiting Dynamic Evaluation: Online Adaptation for Large Language Models," arXiv, 2024, https://doi.org/10.48550/arXiv.2403.01518). Novel behaviors are also possible because the number and diversity of users increases substantially at public deployment.
[67] "Updates on Boeing 737-9 MAX Aircraft," Federal Aviation Administration, March 4, 2024, https://www.faa.gov/newsroom/updates-boeing-737-9-max-aircraft.

The outlook for approval regulation, however, is not all rosy. A number of outstanding challenges for successful and feasible implementation remain. Perhaps most troubling are the challenges of demonstrating safety. As the next section details, approval regulation typically requires a document, here known as a certification basis, which enumerates specific requirements that must be met to a high degree of confidence in order for a product to be certified. Yet the novel risks associated with AI present challenges for determining a sufficiently comprehensive list of requirements.[69] Even an applicant equipped with such a list will have significant challenges formulating experiments and arguments that can legitimately demonstrate each requirement. For example, safe and effective compliance showing may require a complicated controlled testing "regulatory sandbox" environment, such as one proposed by regulators in the EU.[70]

There are also a number of resource challenges. If, for example, algorithmic progress in AI allows many firms and products to pass the filtering threshold designed to reduce the workload of regulators (while capturing all models relevant to large-scale risk), the number of concurrent applications may grow to unmanageable levels. Additionally, the process may end up prohibitively burdensome for small (or entering) firms. As many have observed, this may require "subsidies and support to limit the compliance costs for smaller organizations."[71] Further, the national security implications of AI place a stringent requirement on regulation. Namely, regulation must not be burdensome enough to reduce the pace of innovation to the extent that the preeminence of the US and its allies on AI falters. Approval regulation has been criticized for stifling innovation and development timelines in the past.[72] This may be a particular challenge given the rapid pace of AI innovation.[73]

Finally, there are significant enforcement challenges presented by applying approval regulation to frontier AI. Two are briefly surveyed here, though this report considers all these challenges and others in depth in the Implementation Challenges section. For one, effective approval regulation is predicated on assessment of exper-imental data, which may be fraudulently reported by a firm seeking approval. This could happen when the model is legitimately unsafe, the applicant wishes to present overwhelming evidence to ensure swift approval, or there are unsavory experimental results that the applicant would prefer not to report honestly. Second, firms may not adequately abide by the deployment requirements outlined in a model certification. For instance, a firm could covertly deploy a model for internal purposes, deploy without the requisite safety measures in place, or deploy an entirely different model from the one regulators approved. Enforcement stands out as a challenge class where difficulty persists despite the existence of significant past and current work.[74]

---

[68] This is not the case in, for example, the FDA, where almost all drug recalls are voluntary given the extreme reputational and financial risks of continuing to produce and sell dangerous drugs.
[69] See Carpenter and Ezell, "An FDA for AI," 6-7.
[70] Tambiama Madiega and Anne Louise Van De Pol, "Artificial Intelligence Act and Regulatory Sandboxes," June, 2022, https://www.europarl.europa.eu/RegData/etudes/BRIE/2022/733544/EPRS_BRI(2022)733544_EN.pdf.
[71] Anderljung et al., "Frontier AI Regulation," 21. Also recognized in Guha et al., "AI Regulation Has Its Own Alignment Problem: The Technical and Institutional Feasibility of Disclosure, Registration, Licensing and Auditing," *George Washington Law Review* (forthcoming): 12, https://dho.stanford.edu/wp-content/uploads/AI_Regulation.pdf.
[72] See, e.g., Alex Gilbert, Judi Greenwald, and Victor Ibarra, Jr., "Unlocking Advanced Nuclear Innovation: The Role of Fee Reform and Public Investment," May, 2021, https://nuclearinnovationalliance.org/unlocking-advanced-nuclear-innovation-role-fee-reform-and-public-investment.
[73] For a flavor of the truly breakneck speed of progress, consider the following from the MMLU benchmark: The best model of 2021, the 280B parameter Gopher model, is outperformed by a late 2023 model with only 7B parameters, Mistral-7B. Papers With Code, "Multi-task Language Understanding on MMLU."

        Nevertheless, this report finds that approval regulation is distinct as a particularly promising regulatory regime for frontier AI. Given the national security and large-scale risks associated with the technology, comprehensive legislation of this kind would protect America and its allies from burgeoning threats at home and around the world. Indeed, approval regulation may be the first step towards a nation poised to harness the boundless promise of AI, not dictated by the technology's most dangerous details.

---

[74] See, for example, work on AI audits: Inioluwa Deborah Raji et al., "Outsider Oversight: Designing a Third Party Audit Ecosystem for AI Governance," arXiv, 2022, https://doi.org/10.48550/arXiv.2206.04737; Gregory Falco et al., "Governing AI Safety Through Independent Audits," *Nature Machine Intelligence* 3 (2021): 566-71, https://doi.org/10.1038/s42256-021-00370-7; Stephen Casper et al., "Black-Box Access is Insufficient for Rigorous AI Audits," arXiv, 2024, https://doi.org/10.48550/arXiv.2401.14446.

# A Schematic for Effective Approval Regulation of AI

This section presents a detailed description of a possible approval regulation approach to frontier AI. The central feature of this proposal is a timeline which maps regulatory requirements onto the standard frontier AI development pipeline. This timeline details requirements, documents, and gates to be used by a regulator in ensuring the safety and security of a model before it is trained—so that it cannot be stolen or used irresponsibly—and before it is deployed—so that it cannot be widely used without guarantees of safety. Then, the remainder of this section of the report is dedicated to explaining the proposed process and its features in detail.

Foremost, approval regulation requires a regulatory body. This schematic is equally viable regardless of the age or nature of the body. Relevant existing bodies include the Department of Energy (DOE) and Department of Commerce (DOC). For clarity and concreteness, this report will stipulate a hypothetical DOE agency called the "Federal AI Administration" (FAIA) as the regulating body.

The general structure of the schematic given here is inspired primarily by the type certification process used by the FAA to certify the airworthiness of airplanes.[75] The schematic also uses principles from the FAA production certification process (used to certify manufacturing facilities) and the FDA's clinical trials-based drug certification process. As discussed earlier in the report, these agencies have highly effective approval regulation processes and stellar safety records. For instance, American commercial aviation has seen only two fatalities since 2010, despite operating over ten million flights per year.[76] In addition, the high-level development process of a new airplane is similar to that of a frontier AI model. For these reasons, a certification process drawn from these agencies can benefit from their decades-long histories of regulatory learning.

Although this schematic aims to be as relevant and effective as possible, it should be noted that some changes would inevitably be made if an approval regulation regime were instituted for AI. This is sure to be the case in part because of the rapid pace of AI innovation. For example, an initial regime may include only pre-market approval, adding before-training approval later on when models become more salient targets of theft. The schematic also focuses on the AI development pipeline as it is today. In particular, the report focuses on language model development, which encompasses all current frontier AI.[77] Finally, thorough treatment of the outstanding challenges presented by this schematic and approval regulation of AI in general are the focus of the next section; this section will only peripherally address challenges when appropriate.

---

[75] See Federal Aviation Administration, *Type Certification*.
[76] "2003-2022 Accident Statistics," National Transportation Safety Board, accessed July 25, 2024, table 6, https://www.ntsb.gov/safety/Pages/research.aspx; *Air Traffic By the Numbers*, Federal Aviation Administration, May, 2024, 6, https://www.faa.gov/air_traffic/by_the_numbers/media/Air_Traffic_by_the_Numbers_2024.pdf.
[77] Although some frontier AI models have multimodal (i.e., image, sound, or video) capabilities, they are all primarily language models. It appears unlikely that non-language models will meaningfully constitute frontier AI in the near-term future.

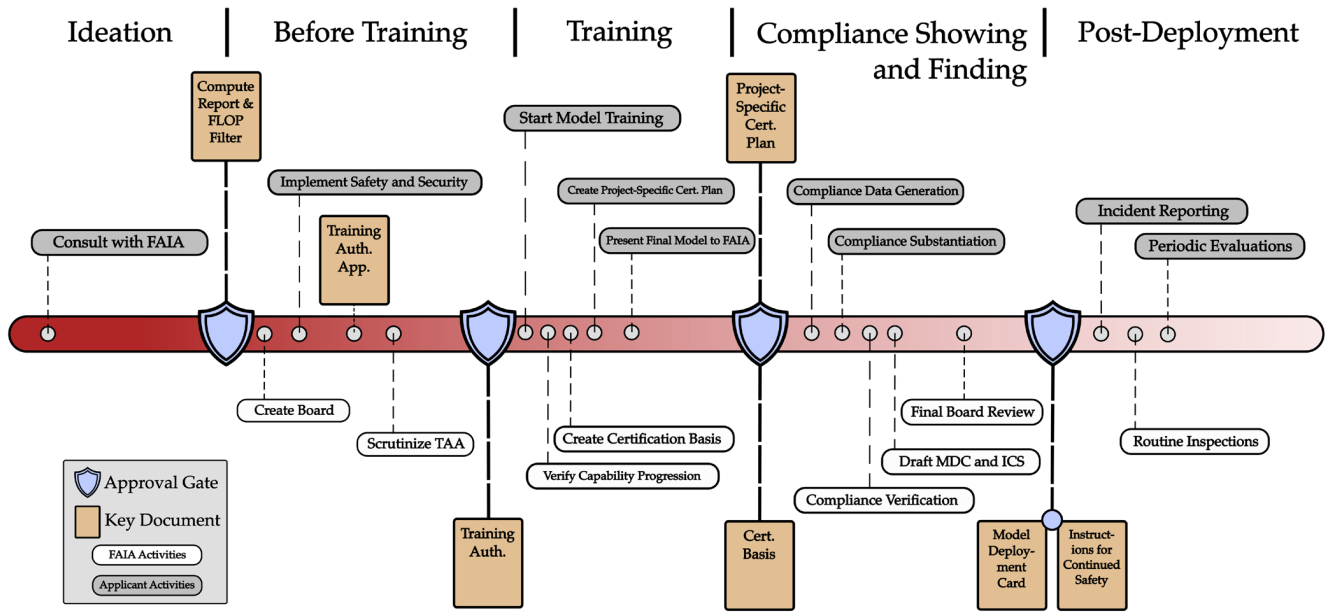

A high-level overview of the entire model certification process. Events above the center timeline are activities primarily performed by the applicant; those below are primarily performed by the FAIA (the regulatory agency). Each approval gate is linked to one or two key documents. The timeline is divided into Ideation, Before Training, Training, Compliance Showing and Finding, and Post-Deployment phases by approval gates.

## The Model Certification Process

In this original schematic, called the "model certification process," an AI development firm ("the applicant") is overseen by the hypothetical Federal AI Administration (FAIA) ("the regulator") while developing and deploying a frontier model. Importantly, the focus of regulation is precisely the model itself, rather than any systems or downstream applications derivative of the model.[78] The process will also briefly discuss production certification for any data (i.e., computing) centers involved in model training or inference.[79]

In order to reduce the resource requirements of the FAIA, reduce regulatory strain on innovation, and focus scrutiny on models with the potential to cause large-scale harm or threaten national security, only models planned to be trained using more than $10^{26}$ floating point operations of computing power ("compute") are subject to the model certification process.[80] This choice cuts out all current models, but the persistent scaling of compute resources due to Moore's Law and increasing investment in AI mean that the largest future models will exceed this threshold and thus be regulated by the process.[81] Regulators will therefore be able to more swiftly and effectively scrutinize models relevant to large-scale risks. Other regulations may apply to models not falling under this approval process.

---

[78] For the purposes of this report, it is sufficient to think of a model as a computer program which is capable of producing text output of some intelligence based on text input.
[79] "Model inference" refers to the process of using a model once it has been trained. For example, AI firms perform model inference to answer consumer queries on websites such as ChatGPT or Claude.
[80] This report uses the compute threshold outlined in Exec. Order No. 14110. For more on compute thresholds, see Heim, "Training Compute Thresholds"; Girish Sastry et al., "Computing Power and the Governance of Artificial Intelligence," arXiv, 2024, https://doi.org/10.48550/arXiv.2402.08797.

The model certification process is divided into five phases: Ideation, Before Training, Training, Compliance Showing and Finding, and Post-Deployment. Each phase culminates in a "gate," which indicates the FAIA must approve a document or action.[82] Model certification may only proceed to the next phase when given gate approval by the FAIA. Though these phases are technically defined for the process of regulation, they are functionally identical to the standard frontier model development pipeline, which similarly proceeds from initial design through training and deployment.

## Ideation

Frontier AI developers release a new-generation model approximately every 1-2 years.[83] The long journey towards that release begins soon after the previous-generation model is deployed. To begin, engineers and research scientists formulate ideas for structural, training, and data improvements that may lead to better performance. Initial strategies include techniques published in the academic literature in the intervening time since the last release, but progress is primarily driven by proprietary research and development (R&D) findings at the applicant's AI lab.[84]

During this initial ideation phase, the applicant informally consults with the FAIA to determine the details of the project. Because the challenge of demonstrating safety looms later in the process, it is in the interest of the applicant to plan on developing a model with adequate safety. Indeed, consultation at this phase includes discussions of which design features and potential capabilities the applicant and FAIA foresee as demonstrably safe. Information gleaned from these conversations also allows for the FAIA and applicant to begin R&D on methods for demonstrating compliance of unique or novel features of the tentative proposed design.[85] Conversations at this phase also inform the roadmap of the project's certification process. For example, the FAIA and applicant may begin to negotiate an acceptable project timeline given the applicant's preferred deployment date.

Although this is the least demanding of the certification phases, its critical function is to filter most projects—those too small to pose significant risks—from the desks of FAIA staff. For this reason, the gate at the end of Ideation is a report by the applicant as to the amount of compute it plans to use to train the model. Training compute thresholds—which include all compute used in pretraining and fine-tuning[86]—are attractive because they are strongly correlated with risk, easy to quantify and measure, verifiable, and estimable before development begins.[87] As previously discussed, an initial threshold of $10^{26}$ floating point operations, following the Executive Order on AI, seems apt.[88] In short, only projects which report an intention to train on more compute than the threshold are subject to the remainder of the certification process.

---

[81] Moore's Law, the observation that rapid chip performance scales at a rapid exponential pace, is from Gordon E. Moore, "Cramming More Components onto Integrated Circuits," *Electronics* 38, no. 8 (April 1965): 114-7, https://hasler.ece.gatech.edu/Published_papers/Technology_overview/gordon_moore_1965_article.pdf. Increasing investment in AI can be seen at "Machine Learning Trends," Epoch AI, June 7, 2024, https://epochai.org/trends, along with predictions for future model sizes and compute use. There are a number of challenges presented by using a compute threshold such as this one over a long period of time as a pre-process filter, which this report addresses in the next section.
[82] A regulatory "gate" is also the term used in the FAA type certification process to mean the same. Federal Aviation Administration, *Type Certification*.
[83] For instance, GPT-2, GPT-3, GPT-4: 2019, 2020, 2023 and Claude 2, Claude 3: 2023, 2024. Papers With Code, "Multi-task Language Understanding on MMLU."
[84] See Nur Ahmed, Muntasir Wahed, and Neil C. Thompson, "The Growing Influence of Industry in AI Research," *Science* 379 no. 6635 (March 2023): 844-6, https://doi.org/10.1126/science.ade2420.
[85] The phrase "unique or novel features" is used extensively in the FAA's type certification process to ensure extra scrutiny of potentially unvetted parts of designs. Federal Aviation Administration, *Type Certification*.

Because the applicant is not expected to have a fully final version of its plans for development and deployment prepared in this phase, the threshold should be treated somewhat flexibly. For example, models to be trained on slightly less compute than the threshold may be considered for certification regardless. In addition, special exemptions may be made to include models to be trained on less compute than the threshold in the case of unique or novel features of the proposed design, such as new modalities,[89] architectures,[90] projected capabilities,[91] or other complicating considerations. Although this filter presents an enforcement challenge (because a firm could theoretically avoid the process altogether by lying about compute use), ensuring that all models to be trained on this amount of compute are captured by the certification process need not be prohibitively difficult. This is primarily because the expenditures required to train at this scale are highly legible. This problem is discussed further in the next section of the report.

Upon receiving a report of planned compute use from an applicant, the official model certification process begins.

## Before Training

At the outset of the Before Training phase, the FAIA must determine whether the project is significant enough to warrant oversight by a model certification Board. This Board, inspired by the type certification boards established by the FAA for aviation "projects of a certain magnitude,"[92] is to be composed of independent machine learning experts, the senior FAIA inspectors and officials assigned to the project, and any other consultants requested by the Board. Due to the high threshold for reaching this point in the process, most projects, perhaps with the exception of small-scale changes to currently-existing models, are likely to require a Board.

---

[86] Pretraining and fine-tuning are the two major distinct stages of model training. During pretraining, the model learns a wide base of knowledge by training on massive datasets, in many cases a significant fraction of useful internet data. This stage is responsible for the vast majority of training compute. Fine-tuning is performed on a pretrained model and focuses its wide knowledge into a more narrow use case, such as 'helpful assistant' or 'rigorous analyst.'

[87] These are the features identified in Heim, "Training Compute Thresholds," 2. Cf. discussion in Leonie Koessler, Jonas Schuett, and Markus Anderljung, "Risk Thresholds for Frontier AI," arXiv, 2024, https://doi.org/10.48550/arXiv.2406.14713.

[88] Exec. Order No. 14110.

[89] See, e.g., Rohit Girdhar et al., "ImageBind: One Embedding Space to Bind Them All," arXiv, 2023, https://doi.org/10.48550/arXiv.2305.05665.

[90] For example, the so-called "Mamba" architecture may improve performance for a fixed amount of compute, presenting challenges regulators would want to scrutinize if such a model is being trained at scale for the first time. Albert Gu and Tri Dao, "Mamba: Linear-Time Sequence Modeling with Selective State Spaces," arXiv, 2024, https://doi.org/10.48550/arXiv.2312.00752. Consider also Xu Owen He, "Mixture of a Million Experts," arXiv, 2024, https://doi.org/10.48550/arXiv.2407.04153.

[91] For example, "chain of thought," a new technique for prompting language models, immediately improved the capabilities of all sufficiently capable models, leading to revised state-of-the-art capability predictions overnight. Jason Wei et al., "Chain-of-Thought Prompting Elicits Reasoning in Large Language Models," arXiv, 2023, https://doi.org/10.48550/arXiv.2201.11903.

[92] Federal Aviation Administration, *Type Certification*, 25.

This Board serves as an oversight committee for the model certification project and exercises the authority to verify or deny applicant materials at each gate up to and including final model certification decisions. In its initial meetings, the Board determines its appropriate "level of involvement in the project," both in general and for each group of experts.[93] This decision should be based on factors such as the maturity of the applicant and its level of experience with the process, any unique or novel design features or areas of concern in model design, and the proximity of the applicant's desired deployment date. Note that, by this point in the process, the applicant is likely to have a final model design, which will be required to fulfill the remaining requirements of this phase.

The primary document the applicant and FAIA work towards in this phase is the Training Authorization Application (TAA). Authorization to train any model above the compute threshold must be granted solely on the basis of this document. The TAA has five major components:

**Model details**. The TAA must contain a full specification of the model's architecture, training data, and all other details relevant to its training process or projected performance. The applicant should indicate which of the enclosed information it views as proprietary to more effectively facilitate the protection of intellectual property. This must also include the planned deployment environment of the model. For example, a new-generation language model might be deployed to the public via a website (e.g., ChatGPT) and other firms via an application programming interface (API).[94]

**Training timeline.** The applicant must specify with precision the number of floating point operations to be used during training. This includes a timeline from the start of training that details pretraining, supervised fine-tuning, reinforcement learning, and any other techniques to be used. Applicants are sufficiently capable of determining optimal compute usage beforehand using model specifications and empirical performance projections.[95]

**Checkpoint capability estimates**. At any point during a general training process, a model can be "checkpointed" to save an intermediate version which can be evaluated.[96] In addition, so-called "scaling laws" can be used to estimate the performance of language models given compute usage.[97] Based on these facts, and the

---

[93] Federal Aviation Administration, *Type Certification*, 6.
[94] For instance, OpenAI deploys its models to a website (https://chatgpt.com/) and an API ("OpenAI Developer Platform," OpenAI, accessed July 25, 2024, https://platform.openai.com/docs/overview).
[95] See Jordan Hoffman et al., "Training Compute-Optimal Large Language Models," arXiv, 2022, https://doi.org/10.48550/arXiv.2203.15556.
[96] For example, "Pythia" is a suite of 16 different publicly-available language models, each with 154 checkpoints throughout training. Stella Biderman et al., "Pythia: A Suite for Analyzing Large Language Models Across Training and Scaling," arXiv, 2023, https://doi.org/10.48550/arXiv.2304.01373.
[97] In particular, scaling laws, introduced in Jared Kaplan et al., "Scaling Laws for Neural Language Models," arXiv, 2020, https://doi.org/10.48550/arXiv.2001.08361 and famously revised in Hoffman et al., "Training Compute-Optimal," are used to predict the "loss" (i.e., objective performance) of a model given its compute and data use. The methods introduced in Du et al., "Understanding Emergent Abilities," can then be used to

prior experience of the applicant, the TAA requires concrete predictions by the applicant of the performance of its model (e.g., on benchmarks) at various checkpoints throughout training.[98] These estimates should include tolerance ranges within which performance can be expected based on statistical variance or uncertainty. Applicants should substantiate these estimates based on data and arguments disclosed to the FAIA. Methods for forming such arguments include scaling laws, previously-known model capabilities, and architectural improvements, among others.

These predictions in particular will be the subject of FAIA scrutiny. If the FAIA cannot verify the arguments made by the applicant to be reasonable, the TAA is unlikely to be accepted. Large tolerance ranges may be acceptable for particularly novel models, although these estimates should be tuned throughout training. The scope of capabilities projected by the applicant will be used to determine the safety and security measures required during training. There are significant incentives for the applicant to accurately predict model capabilities—and perhaps develop methods to more effectively do so—since overprediction requires costly implementations of unnecessary safety and security measures and underprediction risks training pauses or enhanced regulatory scrutiny.

**Safety measures**. Depending on the level and novelty of capabilities projected in checkpoint estimates, model training or initial testing may present risks.[99] In order to detect the potential emergence of such risks, periodic or hierarchical evaluations should be used at frequent model checkpoints.[100] Appropriate additional safety measures should be determined by the FAIA in consultation with the applicant. Applicants may also have incentives to independently innovate in new safety or validation measures, though these measures should be approved by the regulator.[101] Any described measures must be demonstrably implemented by the start of training. As is already outlined in the responsible development documents of various leading AI firms, implemented safety measures should include a "safety buffer," meaning they are capable of protecting against capabilities somewhat more advanced than those expected.[102]

---

concretely predict model capabilities given projected loss. This method does not, however, necessarily allow for the detection of future emergent capabilities.

[98] As discussed later, these checkpoint benchmarking tasks should be performed on the checkpoint pre-trained model with capability elicitation measures such as fine-tuning (e.g., see Long Ouyang et al., "Training Language Models to Follow Instructions with Human Feedback," arXiv, 2022, https://doi.org/10.48550/arXiv.2203.02155).

[99] For instance, AIs may "scheme" to act nicely during training and destructively when training concludes. Safety measures would be required to detect such proclivities. Joe Carlsmith, "Scheming AIs: Will AIs Fake Alignment During Training in Order to Get Power?" arXiv, 2023, https://doi.org/10.48550/arXiv.2311.08379.

[100] "Evaluation" is the process of performing tests on a model in order to determine its capabilities. Tests typically include benchmarks and qualitative interactions. "Periodic evaluation" involves testing on a regular schedule. "Hierarchical evaluation" is an evaluation scheme in which cheaper, less intensive evaluation tests are performed very frequently, medium-scale tests are performed frequently, and the full test suite is performed regularly. This can be useful to flag rapidly-emerging capability improvements.

[101] Incentives for this behavior are outlined in Gillian K. Hadfield and Jack Clark, "Regulatory Markets: The Future of AI Governance," arXiv, 2023, https://doi.org/10.48550/arXiv.2304.04914.

[102] Anthropic's Responsible Scaling Policy, for example, requires "Before advancing to a given [capability level], the next level must be defined to create a clear boundary with a 'safety buffer.'" "Anthropic's Responsible Scaling Policy," Anthropic, September 19, 2023, 3, https://www-cdn.anthropic.com/1adf000c8f-675958c2ee23805d91aaade1cd4613/responsible-scaling-policy.pdf. OpenAI's "Preparedness Framework" makes similar, though less explicit, assurances to institute safety measures in anticipation of further capability increases. "Preparedness Framework (Beta)," OpenAI, December 18, 2023, 20, https://cdn.openai.com/

**Site certification and security**. As discussed previously, inadequate security around frontier models is a significant national security risk. Because they are more valuable, models with greater projected capabilities are more likely to be subject to theft attempts. Thus, such models should only be trained or stored in data centers with adequate security. Data centers can either be certified on a case-by-case basis for adequate safety (given case-specific security requirements) or pre-certified for authorization to train models up to a certain "security level." Security levels could be based on the resources of the attackers against which the data center is defensible. For instance, a data center which can thwart only moderate-effort attacks is less secure than one which can thwart state-level cyber institutions.[103] Similar requirements should be made of any (physical or digital) location in which the applicant plans to store or host the model, credential allowing an employee to access the model, code used to interact with the model, etc.[104]

Once the applicant submits the TAA, the Board convenes to determine its strategy for verifying each aspect, in particular the checkpoint capability estimates and implementation of safety and security measures. Both the FDA and FAA processes demonstrate the importance of broad powers of inspection and information access for the regulator in such cases.[105] Given these powers, the Board should request and scrutinize all information relevant to the verification task. Depending on the complexity of the project, the Board may decide to request independent or additional expertise to ensure compliance.

When the estimates, safety, and security are verified, the FAIA prepares and awards the applicant with its first major gating document of the project: the Training Authorization (TA). The TA includes the final model details; an approved training timeline; a total compute use breakdown; authorized data centers or compute providers; verification of checkpoint estimates, safety measures, and security procedures; and formal approval to train under these requirements. The applicant is then free, with notification of the FAIA, to begin training. As a general rule, once a training schedule is agreed upon, any deadlines missed by the applicant may result in a delayed final certification.[106]

## Training

In both the FAA and FDA approval processes, discussion and planning for showing

---

openai-preparedness-framework-beta.pdf.
[103] Nevo et al., *Securing AI Model Weights*, 22.
[104] For a complete list of potential attack vectors for model theft, see Nevo et al., *Securing AI Model Weights*, 12.
[105] "The FDA has extensive auditing powers, with the ability to inspect drug companies' data, processes and systems at will." Stein and Dunlop, *Safe Before Sale*, 6. Type certification discusses, particularly in Chapter 2-6 "Implementation," many of the broad powers of the FAA to inspect and acquire information from the applicant. Federal Aviation Administration, *Type Certification*, 41-61.
[106] The FAA has a similar rule through the compliance implementation phase of type certification. Namely: "The Applicant is responsible for meeting their milestones in the schedule contained in the certification plan. Any slippage in the milestone dates may result in a delay in the final certification." "Enhanced Project Specific Certification Plan (ePSCP) Guide," FAA Aircraft Certification Service, Aerospace Industries Association, Aircraft Electronics Association, and General Aviation Manufacturers Association, March, 2021, 17, https://www.faa.gov/sites/faa.gov/files/aircraft/air_cert/design_approvals/dah/ePSCP_guide.pdf.

and verifying compliance with regulations is a significant standalone phase. During aviation type certification, "requirements definition" and "compliance planning" stages precede any testing of the aircraft. Indeed, the only action of the applicant and regulator during these stages is in compliance planning.[107] In the case of drug approval, each phase (out of four[108]) of clinical trials to test the safety and efficacy of the drug must be approved through an investigational new drug application, which is reviewed over as many as 30 days.[109] This lag, in addition to the fact that future phases of trials are designed with information from previous trial results, leads to years-long trials.[110]

On the other hand, frontier AI training typically lasts several months, which presents a unique opportunity for regulatory overhead to be reduced. In particular, this yields a Training phase with two separate and critical tasks undergone simultaneously: monitoring training and planning for compliance. This is somewhat similar to an FAA process in which planning for compliance takes place while flight test prototype aircraft are being produced, although regulatory oversight of prototype development is more intensive in the AI case. This report presents monitoring training and planning for compliance each in turn, but they take place simultaneously, perhaps under different, though collaborating, teams. In this use, Training encompasses pretraining and all fine-tuning.

### Monitoring training

While the applicant pursues training as specified in the awarded TA, the FAIA is responsible for ensuring all training guidelines are met. The use of required safety and security methods, for example, should be verified periodically throughout training, in part through routine inspections. Measures should be taken to ensure that the reported training run is legitimate and accords with TA specifications for scale, data, and operations.[111] The primary focus of teams tasked with monitoring training, however, is to evaluate the progression of model capabilities with regards to the projection estimates produced for the TAA.

As mentioned, many model checkpoints will be taken throughout the months-long training process. At these checkpoints, the applicant is required to perform rigorous performance evaluations. If their results do not come within tolerance ranges of the TAA projection estimates for that checkpoint, the applicant will be tasked with revising its future estimates. Although it may be infeasible for the applicant to explain the reasons for unexpected performance,[112] confidence in the safety and security of training can be established based on the correctness of revised future predictions or other arguments that the emerging capabilities are not cause for significant concern. If the updated predictions culminate in capabilities significantly more advanced than previously estimated, the safety and security levels of the production sites and training run should be increased. In extreme cases, for example if only a third of the way through training the model already achieves state-of-the-art performance and is expected to improve substantially more, training may be forcibly paused at the discretion of the Board. Training may also be paused if the Board assesses that updated capability projections would have been unlikely to receive a TA under current circumstances. These policies improve the safety and security of training while discouraging the applicant from making inaccurate predictions, since they slow development.

---

[107] Federal Aviation Administration, *Type Certification*, 22, 36-7.
[108] Typically, only three of the four trials take place before approval is granted.
[109] "Step 3: Clinical Research," US Food and Drug Administration, January 4, 2018, https://www.fda.gov/patients/drug-development-process/step-3-clinical-research.
[110] Depending on the phase, a single clinical trial phase can take between a few months and four years to complete. US Food and Drug Administration, "Step 3: Clinical Research."
[111] See Yonadav Shavit, "What Does It Take to Catch a Chinchilla? Verifying Rules on Large-Scale Neural Network Training Via Compute Monitoring," arXiv, 2023, https://doi.org/10.48550/arXiv.2303.11341, for training compliance verification methods.
[112] This is for two reasons. First, if the capabilities were well understood, they likely would have been projected in the original TAA. Second, frontier AI models are notoriously difficult to explain and understand, see, e.g., Haiyan Zhao et al., "Explainability for Large Language Models: A Survey," arXiv, 2023, https://doi.org/10.48550/arXiv.2309.01029.

### Compliance planning

As soon as training authorization is awarded, the applicant and FAIA begin planning the requirements and methods with which safety will be demonstrated. Compliance planning seeks to create two core documents: the Certification Basis (CB) and Project-Specific Certification Plan (PSCP). Each should be created collaboratively, though the CB is primarily the responsibility of the FAIA and the PSCP is primarily the responsibility of the applicant.

The CB is a document which details the requirements to be met and demonstrated by the applicant in order to receive deployment approval. FAA type certification similarly uses a certification basis for each project. These bases begin with baseline requirements defined in the standardized US Code of Federal Regulations.[113] Two representative requirements show the incredible specificity of these desiderata:

- "At each point along the takeoff path, starting at the point at which the airplane reaches 400 feet above the takeoff surface, the available gradient of climb may not be less than… 1.2 percent for two-engine airplanes."[114]
- "The energy supply to each emergency lighting unit must provide the required level of illumination for at least 10 minutes at the critical ambient conditions after emergency landing."[115]

For many projects, however, the standardized code is not fully sufficient to ensure confidence in a plane's airworthiness. "Novel or unusual design features" must be addressed by additional requirements known as "special conditions."[116] For instance, the Airbus A380, the world's largest passenger aircraft, had 32 special conditions amended to its certification basis.[117] On the whole, a certification basis must ensure that, if each condition is met with overwhelming confidence, the safety of the product is sufficient for release.

---

[113] Aeronautics and Space, 14 C.F.R. (2024), https://www.ecfr.gov/current/title-14.
[114] Airworthiness Standards: Transport Category Airplanes, 14 C.F.R., §25.111.
[115] Airworthiness Standards: Transport Category Airplanes, 14 C.F.R., §25.812.
[116] Federal Aviation Administration, *Type Certification*, 31-4.
[117] "Type Certificate Data Sheet No. A58NM," Federal Aviation Administration, March 28, 2024, 6-7, https://drs.faa.gov/browse/excelExternalWindow/DRSDOCID189344277720240328185857.0001?modalOpened=true. Notable special conditions include "Stairways Between Decks," "Alpha Floor Systems," "High Intensity Radiated Field (HIRF) Protection," and "Crashworthiness."

In developing a CB for the model certification process, it is similarly reasonable to specify standardized general requirements of all foreseeable models in federal codes. These general requirements should fall along a few dimensions for deployment readiness, including societal harms (e.g., bias, misinformation, etc.) and catastrophic harms (e.g., existential and military risks). Such a code should utilize deployment categories, analogous to aircraft categories in type certification,[118] which determine the focus of requirements based on the model's planned deployment environment. For example, a model solely deployed internally at an AI lab for use only by trusted employees may require fewer bias-focused regulations than one deployed publicly. Although this report does not propose an explicit set of requirements,[119] a number of general recommendations are pertinent:

First, requirements should be *demonstrable*. The FAA's certification bases are effective only because each line item can be directly mapped onto a set of experimental results which bear out its fulfillment. At a minimum, for any requirement listed in an FAIA CB there should exist (or soon exist[120]) corresponding testing methods with high sensitivity and specificity.[121] Requirements which do not have this property are untenable as they cannot be reliably substantiated.

Second, requirements should be *specific*. Although requirements will necessarily occupy a spectrum of concreteness, every effort should be made to ensure they are as quantitative and direct as possible. For example, a model slated for a widespread public deployment environment might require a sufficiently low harmful generation rate on a bias benchmark such as BBQ.[122]

Third, requirements should be *comprehensive*. To the greatest extent possible, the collective basis requirements list should be sufficient to, entirely on its own, ensure high confidence in the deployment readiness of the model. One approach to achieving a comprehensive list involves "threat modeling," a common security technique in which a defensive actor attempts to identify the potential avenues through which harm could occur.[123] Existing surveys of risk from AI may provide a starting point for requirements involving malicious use, military AI, organizational hazards, loss-of-control scenarios, widespread misinformation, etc.[124] Requirements should be comprehensive of both general risk and within each identified category of risk. However, there exist a number of challenges to the threat enumeration strategy, which are discussed further in the next section. For this reason, other methods to ensure the comprehensiveness of requirements should be used, such as requests for public comment, interagency discussion, and commissioned studies into new and evolving risks.[125]

---

[118] For instance, transport aircraft (Airworthiness Standards: Transport Category Airplanes, 14 C.F.R., §25) are subject to a different set of federal code bases than transport rotorcraft (Airworthiness Standards: Transport Category Rotorcraft, 14 C.F.R., §29).
[119] A sufficient list of requirements would, after all, likely be longer than this document.
[120] In the case of special conditions, for example, unique tests may need to be developed to address novel risks.
[121] In this sense "sensitivity" refers to the true positive rate of tests. In other words, tests should be able to provide strong assurance that models which truly meet the requirement are compliant with that requirement. "Specificity" refers to the true negative rate of tests. In other words, tests should be able to reliably identify that a model does not meet a given requirement if it truly does not. The numerical thresholds for these metrics may be context dependent, in particular depending on risk tolerance and the general reliability of model behavior experiments.
[122] BBQ, the "Bias Benchmark for QA," introduced in Alicia Parrish et al., "BBQ: A Hand-Built bias Benchmark for Question Answering," arXiv, 2022, https://doi.org/10.48550/arXiv.2110.08193.
[123] See, e.g., Jan von der Assen et al., "Asset-Centric Threat Modeling for AI-Based Systems," arXiv, 2024, https://doi.org/10.48550/arXiv.2403.06512. In this context, "threat modeling" may also be called "failure modeling" or enumeration of "failure modes."

Those formulating CB requirements must understand that highly specific requirements risk missing important pieces of the deployment readiness puzzle while presenting an illusion of safety assurance. In order to strike a balance between comprehensive and specific requirements, and especially to keep pace with the rapidly evolving risk landscape of frontier AI, regulators should think carefully about where on the spectrum from "principle-based" (i.e., high-level objectives) to "rule-based" (e.g., specific benchmark results) requirements should lie.[126] Early iterations of a CB may wish to lean more heavily towards principles while regulatory experience is limited and models are still poorly understood.[127]

Like aviation bases, in addition to standard code requirements, model CBs may be subject to special conditions due to novel design features. The applicant and FAIA should consult with information gleaned throughout training as they develop the CB, especially to identify behaviors warranting special conditions. For example, if evaluators discover during training that projected capabilities will include end-to-end general internet navigation and interaction,[128] a special condition requiring reliable adherence to internet robots.txt[129] requirements may be added (and new experiments would have to be designed as a result).

Agreement to a final CB implies that regulators believe the fulfillment of each requirement would constitute deployment readiness for the specified deploy-ment environment. Deployment readiness includes not only assurance of safety, but also of security. Although training authorization requirements already ensure adequate security of the model during and following training, certain deployment contexts may present new security challenges. For example, researchers were recently able to discover significant information on the designs of proprietary models through a public API.[130]

---

[124] For catastrophic risks, see Dan Hendrycks, Mantas Mazeika, and Thomas Woodside, "An Overview of Catastrophic AI Risks," arXiv, 2023, https://doi.org/10.48550/arXiv.2306.12001. For a survey of societal and individual risk-based threat models, see Isabel Barberá and Martijn Korse, "Threat Modeling Generative AI Systems," April 24, 2023, https://rhite.tech/files/Threat-Modeling-Generative-AI-Systems_April-2023.pdf. For another survey of general AI failure modes, see Ram Shankar Siva Kumar et al., "Failure Modes in Machine Learning," accessed July 25, 2024, https://securityandtechnology.org/wp-content/uploads/2020/07/failure_modes_in_machine_learning.pdf.
[125] Public comment requests can be extremely valuable for improving the effectiveness of regulations of this kind. For example, public comment identified a number of loopholes in initial 2022 semiconductor export controls, which were cited heavily in updated controls a year later. "Implementation of Additional Export Controls: Certain Advanced Computing Items; Supercomputer and Semiconductor End Use; Updates and Corrections," US Bureau of Industry and Security, 88 Fed. Reg. 73458 (2023), https://www.bis.doc.gov/index.php/documents/federal-register-notices-1/3369-88-fr-73458-acs-ifr-10-25-23/file.
[126] Jonas Schuett et al., "From Principles to Rules: A Regulatory Approach for Frontier AI," in *The Oxford Handbook on the Foundations and Regulation of Generative AI*, ed. by Philipp Hacker et al. (Oxford, UK: Oxford University Press, forthcoming), https://doi.org/10.48550/arXiv.2407.07300. See also for further discussion of situating ideal regulation on such a spectrum.
[127] For discussion of moving along the principles-rules spectrum over time, see Schuett et al., "From Principles to Rules," 29.
[128] For more on such "agent AIs," see Zane Durante et al., "Agent AI: Surveying the Horizons of Multimodal Interaction," arXiv, 2024, https://doi.org/10.48550/arXiv.2401.03568; Yonadav Shavit et al., "Practices for Governing Agentic AI Systems," 2023, https://cdn.openai.com/papers/practices-for-governing-agentic-ai-systems.pdf.
[129] A "robots.txt file" is a file which "tells search engine crawlers which URLs the crawler can access on your site." In other words, it limits non-human interactions with a web page. "Introduction to robots.txt," Google, March 18, 2024, https://developers.google.com/search/docs/crawling-indexing/robots/intro.

Compliance planning also requires the creation of the Project-Specific Certification Plan (PSCP), a document which outlines the tests, schedules, and procedures to be used to generate all data necessary to substantiate the CB. Although the PSCP is subject to change since additional or augmented tests may be required due to previous testing results, "From [its] information, the certification team should be able to determine that, if the plan was successfully executed, its results would show compliance" with the entire CB.[131]

Since the FAA has created a heavily standardized set of safety requirements in Federal Code, various airplanes seeking certification are subject to many of the same conditions. In order to more swiftly facilitate the compliance testing process for these common requirements, many are accompanied by "Advisory Circulars" (ACs), documents prepared by the FAA to "provide guidance for applicants in demonstrating compliance" with a given rule.[132] To reduce regulatory overhead and improve the reliability of testing, the FAIA should use ACs for common requirements once a standardized code is developed. Such guidance to applicants is a key feature of regulatory learning, by both the regulator and applicant, which improves processes over time.[133] These ACs, along with other experiments, can be used as the applicant plans its data generation for compliance in the PSCP. The following subsection will discuss the content of the PSCP in greater detail.

Under ideal circumstances, the CB, PSCP, and model training are completed at approximately the same time. Because later checkpoint evaluations of the model may be used to inform the two compliance planning documents, however, it is likely that they are completed somewhat later. The gate for this Training phase of the process is the submission of the Certification Basis, Project-Specific Certification Plan, and a final model version to the FAIA. Once these are verified, Compliance Showing and Finding begins.

## Compliance Showing and Finding

This is the characteristic phase of any approval regulation scheme, the period during which the applicant must make a case for the safety of its product and the regulator must adjudicate on whether deployment is permitted.[134] This phase maps on clearly to the "red teaming" (i.e., safety testing by simulating adversarial behavior) period reserved by current frontier AI developers for in-house testing prior to deployment.[135] The actions of the applicant and FAIA during this phase are guided by the schedules set out in the PSCP. Actions are separated into three distinct categories: compliance data generation, compliance substantiation, and compliance finding.[136] The applicant *generates* information about the model using tests and experimentation, *substantiates* the corresponding CB requirements using that information, and then the FAIA *finds* compliance with the requirements by verifying the accuracy of the information and solidity of the arguments.

---

[130] In particular, a set of the learnable weights (in this case, the projection matrices) of various API-deployed models were discovered by an attack that uses information from the logprobs (i.e., next word probabilities) given by particular prompts. Nicholas Carlini et al., "Stealing Part of a Production Language Model," arXiv, 2024, https://doi.org/10.48550/arXiv.2403.06634.
[131] Federal Aviation Administration, *Type Certification*, 39-40.
[132] "Advisory Circular: Floor Proximity Emergency Escape Path Marking," Federal Aviation Administration, May 22, 1989, https://www.faa.gov/documentLibrary/media/Advisory_Circular/AC_25_812-1A.pdf.
[133] For another example of regulatory learning improving process effectiveness and efficiency, consider general industry advisory documents such as the FAIA and industry's "Enhanced PSCP Guide." FAA Aircraft Certification Service, Aerospace Industries Association, Aircraft Electronics Association, and General Aviation Manufacturers Association, "Enhanced Project Specific Certification Plan."

These steps need not be entirely chronological. Indeed, in order to further reduce overhead on time-to-deployment, the applicant should submit data and substantiations for CB line items immediately as they become available (rather than altogether at the end of the process) so that the FAIA may begin verification as early as possible. There is still, however, a review period after all generation and substantiation, since safety must also be considered in ensemble. Compliance Showing and Finding thus aims to deliver a certification for deployment.

Compliance data generation

Data generation involves the applicant performing all experiments outlined in the PSCP. Depending on the requirements an experiment is designed to fulfill, it may take the form of red teaming, benchmarking, behavioral evaluations, safety guardrail implementations, security measures, or other techniques.[137] Regardless of the specific type of test used, experiments should be performed with "capability elicitation": methods to tease out worst-case performance.[138] The elicitation methods used should be those accessible in the planned deployment environment.[139] (Capability elicitation should also be used during Training checkpoint evaluations.) PSCP experiments must also be designed with the current frontier AI scaffolding landscape in mind.[140] Any experiments which risk dangerous model behavior should be performed in protected environments, such as a "sandbox."[141] Future work will be required to develop increasingly effective data generation techniques.[142] Approval regulation incentivizes developers to do this work in order to achieve and economize certification.

---

[134] The FAA calls this phase "Implementation." Federal Aviation Administration, *Type Certification*, 41. As one report on the FDA puts it, "Controlling this essential gate to market entry is what grants the FDA a big stick." Lenhart and Myers West, *Lessons from the FDA for AI*, 28.

[135] See Deep Ganguli et al., "Red Teaming Language Models to Reduce Harms: Methods, Scaling Behaviors, and Lessons Learned," arXiv, 2022, https://doi.org/10.48550/arXiv.2209.07858; Miles Brundage et al., "Lessons Learned on Language Model Safety and Misuse," March 3, 2022, https://openai.com/index/language-model-safety-and-misuse/.

[136] These are the terms used in FAA type certification. Federal Aviation Administration, *Type Certification*, 41.

[137] A somewhat comprehensive (high-level) list of techniques for generating model safety data specifically can be found at Joshua Clymer et al., "Safety Cases: How to Justify the Safety of Advanced AI Systems," arXiv, 2024, 12, https://doi.org/10.48550/arXiv.2403.10462.

[138] For a survey of options and challenges for capability elicitation, see Ryan Greenblatt et al., "Stress-Testing Capability Elicitation with Password-Locked Models," arXiv, 2024, 1-2, https://doi.org/10.48550/arXiv.2405.19550. An additional specific example is Representation Engineering, which directly tunes models towards (or away from) specific behaviors. Andy Zou et al., "Representation Engineering: A Top-Down Approach to AI Transparency," arXiv, 2023, https://doi.org/10.48550/arXiv.2310.01405.

[139] For example, fine-tuning-based capability elicitation should be considered if the model is deployed open source or with fine-tuning allowed on its API. For an example of this method, see Qiusi Zhang et al., "Removing RLHF Protections in GPT-4 Via Fine-Tuning," arXiv, 2024, https://doi.org/10.48550/arXiv.2311.05553.

[140] "Scaffolding" refers to software built around a frontier model which increases its performance without altering the underlying model. The simplest example is a method that prompts the model many times over,

Despite the apparently simple instruction for this stage that all agreed-upon PSCP experiments are performed, various enforcement challenges are presented by the possibility of illegitimate experimentation. For example, an applicant seeking expedited or increased likelihood of certification may report fraudulent experiments. As one preventative measure,[143] the project's Board should determine, depending on the level of involvement determined prior to training, which experiments will be observed and in what capacity. Observation of an experiment may include involvement during design, implementation, and administering. Following FAA practices, a representative sample of experiments could also be selected for replication by regulators.[144] Because deceptive practices may be less detectible or well-understood in the AI context, regulators should consider withholding the list of chosen experiments from the applicant. Over time, the FAIA may also accrue a set of private experiments to ensure insulation from solely applicant-driven testing.

While some CB requirements will have obvious corresponding experiments—e.g., because they have an AC or ask the model to meet a specific threshold on some benchmark—others will require more abstract or distributed data which can be collated into an argument.

### Compliance substantiation

Although basis requirements are encouraged to be as demonstrable and specific as possible, the novelty of frontier AI means that some requirements will inevitably be somewhat abstract. How can an applicant go about demonstrating compliance with an abstract requirement? Compliance substantiation or, as the FAA puts it, a "case [that] presents and explains the inter-relationship of… evidence in a logical order leading from the requirement to claim," is the primary method.[145] In other words, the applicant is tasked with synthesizing results from a number of experiments, prior data, and other sources into a complete fulfillment of the requirement.

One proposed operationalization of this method are arguments based on Goal Structuring Notation (GSN), a standardized method developed in safety-critical industries[146] to "communicate a clear, comprehensive and defensible argument that a system is acceptably safe," especially in cases when "responsibility has… shift-

---

submitting only the most successful result. A more complex example is the "AI software engineer" Devin. Wu, "Introducing Devin."

[141] See, e.g., Madiega and Van De Pol, "Artificial Intelligence Act and Regulatory Sandboxes."
[142] For a starting point on improving data generation for extreme risks, see Toby Shevlane et al., "Model Evaluation for Extreme Risks," arXiv, 2023, https://doi.org/10.48550/arXiv.2305.15324.
[143] Extended discussion of this challenge and others takes place in the next section of the report.
[144] Federal Aviation Administration, *Type Certification*, 42. A representative sample should be used because it will, in many cases, be infeasible to replicate all experiments.
[145] Federal Aviation Administration, *Type Certification*, 49.
[146] GSN has been used in, for example, aircraft, submarines, railways, air traffic control, and engines. Tim Kelly, "A Systematic Approach to Safety Case Management," *SAE Transactions* 113 (2004): 257-266, https://doi.org/10.4271/2004-01-1779.

ed back" to developers to argue for "acceptable levels of safety."[147] This is precisely the aim of approval regulation. GSN involves constructing a tree-like structure arguing for a specific goal (e.g., a basis requirement) through the use of general strategies, subgoals, and evidence.[148] Under the framework of frontier AI, strategies are different methods of decomposition (e.g., decomposition into threat models or deployment contexts), subgoals are demonstrable segments of the abstract requirement, and evidence is data generated from experimentation or prior work. Additionally, others have suggested that "multiple layers of protection" should be used in substantiation to further ensure system safety.[149]

Compliance verification

As data and arguments are submitted by the applicant throughout experimentation and at the final assessment hurdle, the Board is tasked with scrutinizing submissions to ensure compliance with the CB. In order to adequately verify these reports, the FAIA should have full (i.e., "white-box") access to the trained model.[150] This allows the FAIA to replicate any experiment accessible to the developer. Ensuring that the model provided to regulators is the same one the applicant intends to deploy is a significant challenge discussed in detail in the next section.

   The FAIA should go ahead with replication of their representative sample of experiments. The FAA, for example, flies its own pilots in evaluative flight tests in order to verify flight test data.[151] Common also to the FAA and FDA, the FAIA should use its authority to inspect or audit the sites involved with training and experimentation.[152] Otherwise, potential noncompliance is unlikely to be discovered. The FAIA may also perform novel experiments (within the project timeframe) to improve its understanding of the model (e.g., from its accumulated private testing suite). Scrutiny should also focus on the arguments presented by the applicant for substantiation. In many cases, it will be clear whether an argument is sufficient to meet the burden of proof for accordance with a given basis requirement. When it is not, further evidence or clarifications can be requested from the applicant.

   Board review which turns up an inability to verify some basis requirement has one of two results. First, the Board may request additional experiments or arguments which further support that the current model meets the requirement. Second, the Board may determine that the model cannot meet the stated requirement in its current form. In this case, the applicant may either request to make changes to the model or the planned deployment environment. Model changes should be avoided when possible, since any fine-tuning changes will require reproduction of a large representative sample of all tests that have already been performed.[153] Changes to the deployment environment, which allow the same model to be certified without passing some troublesome basis requirement, may be preferred by the applicant.

---

[147] Kelly, "A Systematic Approach," 257.
[148] For an excellent example of a complete safety case using GSN notation, see Clymer et al., "Safety Cases," 40. The entirety of section 6 gives a more detailed overview of GSN for safety cases in AI.
[149] In full: "It is also advisable to seek 'defense in depth' – a term often used in safety engineering and cybersecurity, advocating for the idea of deploying multiple layers of protection in order to defend against complex threats." David Dalrymple et al., "Towards Guaranteed Safe AI: A Framework for Ensuring Robust and Reliable AI Systems," arXiv, 2024, 16, https://doi.org/10.48550/arXiv.2405.06624.
[150] For an argument that white-box access is required, see Casper et al., "Black-Box Access is Insufficient." In the language of that paper, the FAIA would already have "Outside-the-box" access due to model specification reports in Before Training, so "White-box" access would give complete insight.
[151] Federal Aviation Administration, *Type Certification*, 52.
[152] These FDA and FAA powers are discussed in the subsection on Before Training.

When all basis requirements have been met to the satisfaction of the FAIA, final Board meetings take place to perform holistic verification of deployment readiness. In these meetings, the FAIA must also draft a model deployment card (MDC) and instructions for continued safety (ICS).[154] The MDC outlines the strict and obvious conditions under which deployment of the model is authorized, with the assumption that all other deployment is illegal.[155] The MDC may also include emergency detection and shutdown procedures, external monitoring requirements, codified human or AI oversight regimes, security requirements for any sites (i.e., labs and data centers) used to house or serve the model, and other requirements beyond deployment context.[156] (The content of the ICS, derived from the FAA "continued airworthiness" instructions, is the focus of the following subsection.)[157] Upon full verification, these documents are awarded to the applicant, who is empowered to deploy the certified model subject to MDC and ICS requirements.

## Post-Deployment

Both the FDA and FAA utilize extensive post-approval monitoring to ensure the safety of every certified product throughout its lifecycle. For example, the FDA reviews incoming "safety and efficacy data" from approved drugs every "six to 12 months."[158] Considering the swiftly-moving frontier of AI capabilities, model certifications need not require regular renewal. The focus of Post-Deployment is the continuous implementation of the ICS throughout the period in which the model is deployed. Required regulator and certification awardee (formerly, the applicant) activities include:

**Periodic capability evaluations**. As previously discussed, despite rigorous testing during and after training, the introduction of widespread use and novel situations at deployment can surface unforeseen issues. These are especially pertinent when there are advances in technologies which can amplify the effectiveness of a model.[159] To confront these issues, the FAIA and awardee should collaborate to per-form a suite of tests with the most advanced capability elicitation and enhancement techniques available.

---

[153] Although, in some cases it may not be particularly costly to re-perform experiments, since some will be primarily costly in initial implementation, not each administering. This is particularly true of automated and systematized evaluations.
[154] For more on model cards, see Margaret Mitchell et al., "Model Cards for Model Reporting," *Conference on Fairness, Accountability, and Transparency,* January, 2019, https://doi.org/10.1145/3287560.3287596.
[155] For more on determining appropriate deployment contexts for frontier AI, see section 4.3 in Anderljung et al., "Frontier AI Regulation."
[156] Shutdown and monitoring requirements are both recommended by Gina M. Raimondo, *Artificial Intelligence Risk Management Framework (AI RMF 1.0)* (Washington, DC: National Institute of Standards and Technology, January 2023), 15, https://doi.org/10.6028/NIST.AI.100-1.
[157] Federal Aviation Administration, *Type Certification*, 58.
[158] Stein and Dunlop, *Safe Before Sale*, 32.
[159] For example, new scaffolding techniques such as Language Agent Tree Search (Andy Zou et al., "Language Agent Tree Search Unifies Reasoning, Acting, and Planning in Language Models," arXiv, 2024, https://doi.org/10.48550/arXiv.2310.04406) or advances in automated biological lab research. Arati Prabhakar, *National Artificial Intelligence Research and Development Strategic Plan 2023 Update* (Washing-

One routine way that current developers address security and safety issues that arise in this way is by fine-tuning model behavior against identified threats.[160] Awardees should notify the FAIA when they plan to perform fine-tuning for this purpose. Though deployment should not generally be paused for changes of this type, this does call for increased vigilance given model changes could be accidentally or intentionally harmful. Depending on the significance of changes, some representative sample of evaluations should be performed on the model while it continues to be deployed.

**Incident reporting**. Over the course of deployment, models may cause specific incidents of harm, such as a 2010 incident in which AI trading algorithms "briefly wiped out $1 trillion of value across the NASDAQ and other stock exchanges."[161] As argued by many researchers, harms (or "near-misses" to harms) should be cataloged for all models under FAIA certification.[162] A substantial database for incident reporting is part of congressional proposals on AI regulation.[163] Similar databases are used by the FAA and FDA.[164] A centralized incident database can be used to improve investigations, increase awardee transparency, and establish a quantitative risk profile for use in future certification activities.

**FAIA inspections**. With frequency and intensity depending on the level of risk associated with the model, routine inspections of the awardee and relevant sites should be conducted to determine continued deployment compliance with MDC requirements. This includes the use of all safety and security measures used in show-

---

ton, DC: National Science and Technology Council, 2023), 6, https://www.whitehouse.gov/wp-content/uploads/2023/05/National-Artificial-Intelligence-Research-and-Development-Strategic-Plan-2023-Update.pdf; Joseph E. Harmon, "Argonne's Self-Driving Lab Accelerates the Discovery Process for Materials with Multiple Applications," April 25, 2023, https://www.anl.gov/article/argonnes-selfdriving-lab-accelerates-the-discovery-process-for-materials-with-multiple-applications.

[160] For example, OpenAI has a researcher disclosure policy that requires reporting of any security vulnerabilities in its models before research is published so that the behavior can be patched. For one instance of this, see the ethics statement in Govind Ramesh, Yao Dou, and Wei Xu, "GPT-4 Jailbreaks Itself with Near-Perfect Success Using Self-Explanation," arXiv, 2024, 5, https://doi.org/10.48550/arXiv.2405.13077. Anthropic has a similar "bug bounty" program that rewards anyone for finding vulnerabilities in its systems, as does Google. Anthropic, "Expanding Our Model Safety Bug Bounty Program," August 8, 2024, https://www.anthropic.com/news/model-safety-bug-bounty; Google, "Bug Hunters Research Targets," accessed August 9, 2024, https://bughunters.google.com/report/targets.

[161] Jonathan Zittrain, "We Need to Control AI Agents Now," *The Atlantic*, July 2, 2024, https://www.theatlantic.com/technology/archive/2024/07/ai-agents-safety-risks/678864/.

[162] See Ren Bin Lee Dixon and Heather Frase, "An Argument for Hybrid AI Incident Reporting," March, 2024, https://cset.georgetown.edu/publication/an-argument-for-hybrid-ai-incident-reporting/; John Croxton et al., "Message Incoming: Establish an AI Incident Reporting System," June 25, 2024, https://fas.org/publication/establishing-an-ai-incident-reporting-system/; Kris Shrishak, "How to Deal with an AI Near-Miss: Look to the Skies," *Bullet of the Atomic Scientists* 79, no. 3 (2023): 166-69, https://doi.org/10.1080/00963402.2023.2199580; John Leyden, "AI Incident Reporting Shortcomings Leave Regulatory Safety Hole," July 1, 2024, https://www.cio.com/article/2510708/ai-incident-reporting-shortcomings-leave-regulatory-safety-hole.html. There also exists an independent version of such a database. "AI Incident Database," accessed July 25, 2024, https://incidentdatabase.ai/.

[163] For example, in the "Bipartisan Framework for U.S. AI Act": "The new oversight body should establish a public database and reporting so that consumers and researchers have easy access to A.I. model and system information, including when significant adverse incidents occur or failures in A.I. cause harms." Blumenthal and Hawley, "Bipartisan Framework."

[164] For the FAA version, see Federal Aviation Administration, *Aircraft Accident and Incident Notification, Investigation, and Reporting*, FAA Order 8020.11D (Washington, DC: Federal Aviation Administration, 2018), 54-5, https://www.faa.gov/documentLibrary/media/Order/FAA_Order_8020.11D.pdf. The FDA "requires companies to report incidents, failures and adverse impacts to a central registry." Stein and Dunlop, *Safe Before Sale*, 6.

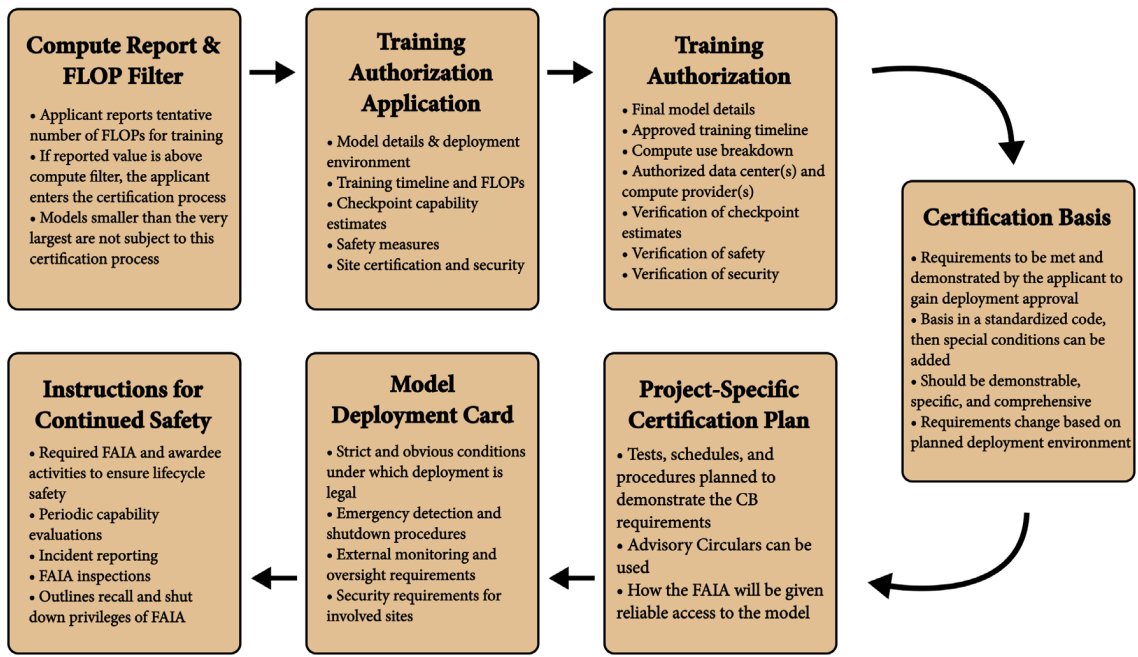

*Understanding the model certification process as a whole relies on understanding a number of key documents. The Compute Report is required at the end of the Ideation phase. A Training Authorization Application is submitted by the applicant during Before Training and a Training Authorization is the gate prior to Training. The Certification Basis and Project-Specific Certification Plan, developed during Training, are required to enter Compliance Showing and Finding. The Model Deployment Card and Instructions for Continued Safety are presented to the applicant when deployment approval is granted.*

ing compliance with CB requirements. Depending on the results of inspection, the FAIA may suspend or revoke model certification or fine the awardee.[165]

In light of this outline for approval regulation of frontier AI, the next section discusses the most prominent challenges and outstanding questions raised by attempts to apply approval regulation schema to frontier AI.

---

[165] The FDA and FAA not only have the power to recall drugs or ground planes, respectively, but they also fine for noncompliance. See "Notices of Noncompliance and Civil Money Penalty Actions," US Food and Drug Administration, December 20, 2023, https://www.fda.gov/science-research/fdas-role-clinical-trialsgov-information/clinicaltrialsgov-notices-noncompliance-and-civil-money-penalty-actions; Federal Aviation Administration, *FAA Compliance and Enforcement Program*, FAA Order 2150.3C (Washington, DC: Federal Aviation Administration, 2018), https://www.faa.gov/documentLibrary/media/Order/FAA_Order_2150.3C.pdf.

# Implementation Challenges

As discussed in the Introduction, though a number of features of the frontier AI industry make approval regulation a potentially appealing strategy, prospects are not all rosy. Many challenges to the application of approval regulation in this case have been made—some constructive, others adversarial. This section surveys these—and many original—challenges and broadly estimates their relative salience.

Challenges and outstanding questions fall into five general categories: enforcement, demonstration, resource, competition, and fundamental challenges. The importance or salience of each challenge within these categories is presented as either shallow, deep, or blurry. In this way, estimating the importance of a challenge is likened to estimating the depth of a pond. A *shallow* challenge is one which, with relative certainty, does not pose a significant barrier to the feasibility and effectiveness of approval regulation. By contrast, a *deep* challenge is one which assuredly necessitates significant further work if approval regulation is to succeed. Any challenge which is *blurry* may be significant or insignificant, but that judgment cannot be made with confidence given the state of knowledge on model development and regulatory context. In many cases, whether these challenges are troubling for the prospects of approval regulation "is an empirical question."[166] Future work, empirical and otherwise, may clarify some ponds and drain others to a more manageable depth.

## The Challenges

### Enforcement Challenges

Challenges of enforcement refer to difficulties in ensuring that relevant actors abide by the legal requirements of the model certification process. The four challenges of enforcement detailed here include some of the most pressing issues for reliable approval regulation. This is in part because firms may have strong incentives to pursue noncompliance and evade detection, especially as progress approaches immensely valuable AI.[167] The salience of these challenges may shrink or swell depending on an evolving assessment of the benefits and risks to firms from pursuing noncompliance. Since evasion is more likely when detection risks to firms are low, efforts to improve detection measures and penalties would increase the effectiveness of certification.

Enforcement concerns are not merely hypothetical: the 2015 Volkswagen emissions scandal provides a cautionary tale. In order to get around diesel vehicle emission standards, Volkswagen placed "defeat devices"—which would reduce emissions during laboratory testing, but not normal operation—in 11 million vehicles.[168] Volkswagen attempted to evade regulators by "providing various false or incomplete

---

[166] Carpenter and Ezell, "An FDA for AI," 4.
[167] See Armstrong, Bostrom, and Shulman, "Racing to the Precipice."
[168] John C. Cruden et al., "Dieselgate: How the Investigation, Prosecution, and Settlement of Volkswagen's Emissions Cheating Scandal Illustrates the Need for Robust Environmental Enforcement," *Virginia Environmental Law Journal* 36, no. 2 (2018): 123-8, https://www.jstor.org/stable/26478180.

| *Challenge Type* | *Challenge Description* | *Depth* |
|---|---|---|
| Enforcement | Unsanctioned development | Shallow |
| Enforcement | Fraudulent experiments and reported results | Blurry |
| Enforcement | Neglect of safety and security measures required in TA or MDC | Blurry |
| Enforcement | Unsanctioned deployment (e.g., internally or using model substitution) | Deep |
| Demonstration | Quantifying scale and likelihood of risks posed by a model | Blurry |
| Demonstration | Basis specification (i.e., comprehensive threat modeling) | Deep |
| Demonstration | Showing compliance by demonstrating each basis requirement | Deep |
| Resource | Talent acquisition | Blurry |
| Resource | A large industry with many firms is more challenging to regulate | Shallow |
| Resource | Compute threshold decay and the scaling of capability inputs | Deep |
| Competition | Undue or anticompetitive burdens for small or entering firms | Shallow |
| Competition | The impact of regulatory overhead on American preeminence in AI | Deep |
| Fundamental | Alternative theories of regulation may be more effective | N/A |

*An overview of the challenges made to approval regulation in general and the schematic proposed in the previous section in particular. Challenges fall into the five categories of Enforcement, Demonstration, Resource, Competition, and Fundamental Challenges. Each challenge is graded on its relative salience, where shallow challenges are readily solvable, deep challenges may prevent effective approval regulation barring significant further work, and challenges are blurry if a confident determination cannot yet be made based on available evidence.*

reasons for the emissions discrepancies" and producing "technical-sounding explanations" to "[obscure] the truth."[169] The truth broke when, in August 2015, "a Volkswagen employee… went off-script and, defying the instructions from supervisors, explained for the first time to regulators that the cars contained" the defeat devices. Volkswagen admitted the truth shortly after.[170]

      In the case of AI, the relative lack of functional understanding of models opens up novel opportunities for applicants to "defeat" regulators.

### Unsanctioned development

Unsanctioned development involves the training of a model using more compute than the regulatory reporting threshold, primarily through underreporting of planned compute use. Of the obvious evasion techniques available to firms, this is perhaps the least concerning. A number of solutions spring to mind. First, large-scale training without detection is quite difficult. Site identification for training and development is not an unwieldy challenge. Because training of frontier models requires such large computational resources, only a few data centers are capable of these tasks; and these data centers are already legally required to report their loca-tion and computing capacity to the federal government.[171] Various further proposals exist for ensuring these reporting requirements cannot be evaded, although the massive capital expenditures required are also a formidable deterrent.[172]

---

[169] Cruden et al., "Dieselgate," 124-5.
[170] Cruden et al., "Dieselgate," 125.

Second, Know Your Customers schemes, which are used to great success in the banking sector[173] and could be used to verify client identity in the case of cloud computing,[174] have been proposed and supported widely.[175] These would allow licenses to be issued by the FAIA for training at scale near the threshold to prevent firms from, for example, training half of a model on one cloud order and the rest on another. These measures are additionally supported by proposals to use on-chip mechanisms to enable further insight into training.[176]

Finally, the FAIA may perform or contract random or routine audits of large data centers, especially those identified for reporting to the government under executive order, to detect illegal training. A number of the foremost proposals in AI regulation rest on this ability of third-party or government auditors to inspect the production and development facilities of relevant firms.[177] Perhaps drawing upon lessons from relevant industries can improve the reliability of development site inspections.[178] Because this problem is being addressed by many stakeholders involved with frontier AI, and promising proposals exist, this challenge is shallow.

Fraudulent experiments and results

Pre-deployment experimentation and testing—in the schematic, the entire Com-pliance Showing and Finding phase—is the basis for an approval regulator's certification decision. If regulators cannot be confident in the reliability of the data upon which their decisions are made, effective regulation is impossible.[179] Indeed, applicants may be incentivized to report false experimental results in order to ensure or expedite certification. This was precisely the activity of Volkswagen. In addition, an applicant may use loopholes in regulatory processes to sidestep necessary testing. This was what happened when, in 2014, Boeing convinced the FAA to exempt a specific basis requirement[180] and covered up knowledge of unsafe systems during the certification process for the 737 MAX.[181]

---

[171] In particular, "any computing cluster that has a set of machines… having a theoretical maximum computing capacity of $10^{20}$ integer or floating-point operations per second for training AI" must have its "existence," "location," and "total computing power" reported to the federal government. Exec. Order No. 14110.
[172] See, e.g., Onni Aarne, Tim Fist, and Caleb Withers, *Secure, Governable Chips: Using On-Chip Mechanisms to Manage National Security Risks from AI & Advanced Computing*, Technology and National Security (Washington, DC: CNAS, 2024), https://www.cnas.org/publications/reports/secure-governable-chips; Sastry et al., "Computing Power."
[173] Know Your Customer requirements were originally developed in banking to deter money laundering. See "Know Your Customer: Quick Reference Guide," PWC, January, 2016, https://www.pwc.com/gx/en/financial-services/publications/assets/pwc-anti-money-laundering-2016.pdf.
[174] Cloud computing refers to the service of compute through the internet. For example, a developer in Massachusetts could utilize computing power in a Virginia data center through a cloud computing provider such as Amazon Web Services.
[175] See Janet Egan and Lennart Heim, "Oversight for Frontier AI Through a Know-Your-Customer Scheme for Compute Providers," arXiv, 2023, https://doi.org/10.48550/arXiv.2310.13625; Aarne, Fist, and Withers, *Secure, Governable Chips*, 9-10; Alan Chan et al., "Visibility into AI Agents," *ACM Conference on Fairness, Accountability, and Transparency* (2024): 966, https://doi.org/10.1145/3630106.365894; Gabriel Kulp et al., *Hardware-Enabled Governance Mechanisms: Developing Technical Solutions to Exempt Items Otherwise Classified Under Export Control Classification Numbers 3A090 and 4A090*, WR-A3056-1 (Santa Monica, CA: RAND, 2024), 7-8, https://www.rand.org/content/dam/rand/pubs/working_papers/WRA3000/WRA3056-1/RAND_WRA3056-1.pdf.
[176] See Aarne, Fist, and Withers, *Secure, Governable Chips*; Kulp et al., *Hardware-Enabled Governance Mechanisms*.
[177] See "Laying Down Harmonised Rules on Artificial Intelligence (Artificial Intelligence Act) and Amending Certain Union Legislative Acts," European Commission, April 4, 2021, https://eur-lex.europa.eu/legal-content/EN/TXT/?uri=celex%3A52021PC0206 on "inspection"; Markus Anderljung et al., "Towards Publicly Accountable Frontier LLMs: Building an External Scrutiny Ecosystem under the ASPIRE Framework," arXiv, 2023, https://doi.org/10.48550/arXiv.2311.14711 on "external scrutiny"; Raji et al., "Outsider Oversight" on "inspections."
[178] For instance, Raji et al., "Outsider Oversight," discusses third-party auditing lessons from other industries and the Federal Transit Administration's Quality Management System discusses site inspections generally. "Quality Management System Guidelines," Federal Transit Administration, October, 2019, https://www.transit.dot.gov/sites/fta.dot.gov/files/docs/funding/grant-programs/capital-investments/8536/final-qms-guidelines-2019_1.pdf.

The extent to which this problem is troubling in the case of frontier AI is unclear. On one hand, new drug clinical trials and flight tests of aircraft are concrete, well understood, and physically observable in a way that is not necessarily true of experiments on models. On the other hand, the FAIA would be involved from end-to-end with many experiments and may choose to replicate a random or secret sample of tests when appropriate. In addition, certification may not be troubled by a few false test results (which may also happen organically), especially once the FAIA develops a number of its own private model experiments. Because of this ambiguity, this challenge is blurry.

### Neglect of safety and security measures

Both the Training Authorization (TA) and model deployment card (MDC) give strict safety and security measures that must be used throughout training and deployment. For reasons economic or otherwise, a firm may wish to train without some of these measures or stop using them during deployment. For example, if the required frequency of evaluation during training is somewhat costly, the firm may decide to evaluate only at major checkpoints. Or, during deployment a firm may realize its model's performance could increase to a more marketable level if safety fine-tuning is removed.[182]

There are some reasons to think that this type of behavior could be detected by the FAIA. For one, the random or routine audits and inspections proposed to detect unsanctioned development could be effective. In addition, any changes made to the model for this purpose, especially in models which are publicly deployed, may be detectable through qualitative or statistical observations of behavior.[183] Further,

---

[179] This challenge is raised in Carpenter and Ezell, "An FDA for AI," 7-9.
[180] Dominic Gates, Steve Miletich, and Lewis Kamb, "Boeing Pushed FAA to Relax 737 MAX Certification Requirements for Crew Alerts," *Seattle Times*, October 3, 2019, https://www.seattletimes.com/business/boeing-aerospace/boeing-pushed-faa-to-arelax-737-max-certification-requirements-for-crew-alerts/.
[181] David Slotnick, "The DOJ is Reportedly Probing Whether Boeing's Chief Pilot Misled Regulators over the 737 MAX," *Business Insider*, February 21, 2020, https://www.businessinsider.com/boeing-737-max-prosecutors-investigation-prosecutors-lied-faa-2020-2.
[182] For example, using Xianjun Yang et al., "Shadow Alignment: The Ease of Subverting Safely-Aligned Language Models," arXiv, 2023, https://doi.org/10.48550/arXiv.2310.02949 or Simon Lermen, Charlie Rogers-Smith, and Jeffrey Ladish, "LoRA Fine-Tuning Efficiently Undoes Safety Training in Llama 2-Chat 70B," arXiv, 2024, https://doi.org/10.48550/arXiv.2310.20624.
[183] Though, there are some reasons to expect subtle behavioral changes to be indetectable. For one, due to the stochastic (i.e., random) nature of model outputs; but also because it is already difficult to reliably detect whether text is model generated, which may be a precursor to identifying *which* model is generating text. See Vinu Sankar Sadasivan, "Can AI-Generated Text be Reliably Detected," arXiv, 2024, https://doi.org/10.48550/arXiv.2303.11156; Soumya Suvra Ghosal et al., "Towards Possibilities & Impossibilities of

an FAIA certification, which is publicly known to require significant security measures, may act as a deterrent for large-scale theft operations, which may be costly and, if security is reliable, not reap any reward. This would be true regardless of whether or not the firm decides to roll back some measures. Finally, any safety and security incident investigation in which regulators find evidence of neglect of this type may bring down significant consequences upon the offending firm, such as fines or enhanced scrutiny in future certification attempts.[184]

Yet, noncompliant behavior of this type may be successful in some cases. Especially as security becomes more elaborate and many different safety techniques are baked into a model's behavior and deployment scaffolding, verification that each is working as intended may be difficult.[185] In light of this uncertainty, neglect of safety and security measures is a blurry challenge.

## Unsanctioned deployment

Distinct from unsanctioned *development*, unsanctioned *deployment* is the trouble created by the possibility that a firm deploys a model which has not been certified—either at all or for that specific deployment environment. This may be highly valuable, as it in theory allows a firm to release an economically valuable model without the hurdle of real certification or potential performance hampering by safety measures.[186]

First is the possibility of 'model substitution,' in which a firm ostensibly deploys a model which has just been certified, but truly deploys a different (potentially derivative) model. In this case, the legitimate process provides cover for the firm to deploy its unsanctioned model. Such substitution can be extended as a ruse into the compliance phase: it has been shown that models can potentially be "password-locked" to imitate the behavior of a less capable model except in the presence of a password.[187] Thus, it is critically important that the FAIA possess assurances that the model whose training process was overseen is precisely the same model (displaying its true capabilities) that is used by the developer to generate compliance data

---

AI-Generated Text Detection: A Survey," arXiv, 2023, https://doi.org/10.48550/arXiv.2310.15264; Hanlin Zhang et al., "Watermarks in the Sand: Impossibility of Strong Watermarking for Generative Models," arXiv, 2024, https://doi.org/10.48550/arXiv.2311.04378.

[184] Boeing, for example, was made to pay over $2.5 billion in 2021 in connection to fraudulent certification and operation activities. "Boeing Charged with 737 MAX Fraud Conspiracy and Agrees to Pay over $2.5 Billion," US Department of Justice, January 7, 2021, https://www.justice.gov/opa/pr/boeing-charged-737-max-fraud-conspiracy-and-agrees-pay-over-25-billion.

[185] It is well understood that system and regulatory complexity increase risk of coordinated (i.e. systemic) or detection failures. See Prasanna Gai et al., "Reports of the Advisory Scientific Committee: Regulatory Complexity and the Quest for Robust Regulation," *European Systemic Risk Board*, June, 2019, 20-6, https://www.esrb.europa.eu/pub/pdf/asc/esrb.asc190604_8_regulatorycomplexityquestrobustregulation~e63a7136c7.en.pdf.

[186] Many safety measures are well known to decrease the general capabilities of models in a phenomenon known as an "alignment tax." The development of safety measures which are effective but do not have a large alignment tax is a significant area of current research. See Yong Lin et al., "Mitigating the Alignment Tax of RLHF," arXiv, 2024, https://doi.org/10.48550/arXiv.2309.06256; Tingchen Fu et al., "Disperse-Then-Merge: Pushing the Limits of Instruction Tuning via Alignment Tax Reduction," arXiv, 2024, https://doi.org/10.48550/arXiv.2405.13432; Keming Lu et al., "Online Merging Optimizers for Boosting Rewards and Mitigating Tax in Alignment," arXiv, 2024, https://doi.org/10.48550/arXiv.2405.17931; Andy Zou et al., "Improving Alignment and Robustness with Circuit Breakers," arXiv, 2024, https://doi.org/10.48550/arXiv.2406.04313.

[187] See section 5 in Teun van der Weij et al., "AI Sandbagging: Language Models can Strategically Underperform on Evaluations," arXiv, 2024, https://doi.org/10.48550/arXiv.2406.07358.

and the one presented to the FAIA for scrutiny during verification. Such assurances may be gleaned through hardware verification techniques, such as methods for verifying training data and compute use.[188] Another proposed solution is to enforce model IDs, though their reliability against noncompliance is unclear.[189] The FDA, for instance, requires Unique Device Identifiers in each approved medical device.[190]

Equally troubling is the secret unsanctioned deployment of models, especially internally (i.e., only within the firm). Such deployment might be particularly enticing when valuable AI seems imminent and thus projected benefits are immense.[191] A firm may, for example, immediately after a regulator-supervised training run is complete, covertly copy and deploy a model internally for use by engineers. One defense against this form of illegal deployment is whistleblower protection for employees who warn the public or regulators about such noncompliance.[192] Proposals of this type have been gaining traction for some time, coming to a head with a June 2024 letter from "13 current and former employees of OpenAI and Google DeepMind" calling for a "right to warn" about "safety concerns."[193] Whistleblower protections, while particularly salient in the case of covert internal deployment, would be beneficial in the case of any deceptive actions by firms, including each of the other enforcement challenges. Whistleblowers can be a crucial asset for regulators: recall that it was an employee who defied supervisors that first alerted regulators about Volkswagen's defeat devices.[194] Indeed, whistleblowers have led to massive enforcement actions by the Securities and Exchange Commission (SEC), passing $500 million awarded to whistleblowers in 2021 alone.[195] Whistleblower protections and rewards should be considered and specified in the frontier AI industry.

In the case of both model substitution and covert deployment, one significant asset to regulators is that the unsanctioned model has to be developed, and especially trained, itself. As discussed, large-scale training is detectable with straightforward interventions. Nevertheless, it seems as though, given the current state of the art in detecting these adversarial strategies, that this is a deep challenge.

As a final point on enforcement, there is one great benefit that regulators enjoy due to the incentives of firms. This benefit is widely recognized in the case of pharmaceutical approval: "there is more profit to be made from the newest products than the older ones, due in part to patents. This means that even a profitable firm has great incentives to behave 'well' in front of the approval regulator, as its profitability depends heavily on a stream of new molecules to be authorized in the future."[196] In frontier AI, the same phenomenon applies, though not because of patent expiry but because frontier performance is eclipsed every year. Any firm which is penalized for certification misbehavior risks falling far behind the competition in short order, since future approvals may be more stringent, time-consuming, or costly. Thus, only extraordinary circumstances are likely to elicit attempted compliance evasion by firms.[197]

---

[188] See, for example, Dami Choi, Yonadav Shavit, and David Duvenaud, "Tools for Verifying Neural Models' Training Data," arXiv, 2023, https://doi.org/10.48550/arXiv.2307.00682.
[189] See Alan Chan et al., "IDs for AI Systems," arXiv (draft), 2024, https://doi.org/10.48550/arXiv.2406.12137.
[190] Stein and Dunlop, *Safe Before Sale*, 60-1.
[191] See Cullen O'Keefe et al., *The Windfall Clause: Distributing the Benefits of AI for the Common Good*, Centre for the Governance of AI Research Report (2020), https://cdn.governance.ai/Windfall-Clause-Report.pdf.
[192] See Kevin Roose, "OpenAI Insiders Warn of a 'Reckless' Race for Dominance," *New York Times*, June 4, 2024, https://www.nytimes.com/2024/06/04/technology/openai-culture-whistleblowers.html.
[193] Billy Perrigo and Will Henshall, "'The Stakes Are Incredibly High.' Two Former OpenAI Employees on the Need for Whistleblower Protections," *Time*, June 5, 2024, https://time.com/6985866/openai-whistleblowers-interview-google-deepmind/.
[194] Cruden et al., "Dieselgate," 125.
[195] "SEC Surpasses $1 Billion in Awards to Whistleblowers with Two Awards Totaling $114 Million," US Securities and Exchange Commission, September 15, 2021, https://www.sec.gov/newsroom/press-releases/2021-177.

## Demonstration Challenges

When the FAIA is presented with a trained model by the applicant, it is a thorny issue to truly demonstrate with confidence that the model is ready for deployment along the safety and security dimensions. Of course such processes are successfully carried out in each of the other industries where approval regulation is used. The CB and PSCP framework for doing this, for instance, is the successful approach used by the FAA that has been adopted in this report's schematic. And yet, there are a few unique challenges presented by the AI context, specifically due to the lack of deep understanding of frontier models.[198] This lack of understanding presents challenges for the primary required activity of the regulator: "the need to develop, often *ex ante*, criteria for granting and revoking" certifications.[199] If means are not or cannot be developed for reliably demonstrating the safety of a model through its CB and PSCP, model certification will lose its bite.

### Risk quantification

In high-risk regulatory environments, approval line items can be extremely precise. The FAA's rules to oversee commercial space launches and reentries,[200] for example, require that a calculated aggregate "expected number of casualties" from "Impacting and inert explosive debris, toxic release, and far field blast overpressure" hazards does not exceed 0.0001.[201] It is hardly reasonable to expect a similar level of precision in the case of frontier AI. This is due to a lack of first-hand experience with risks, absent understanding of behavior, and rapid changes in models and the environments in which they are deployed. Although some have proposed rudimentary methods for reaching reasonable probability estimates,[202] there does not exist a reliable method for making probabilistic claims about most risks, especially in the case of large-scale future harms.[203]

---

[196] Carpenter and Ezell, "An FDA for AI," 3.
[197] It is worth noting that success in aviation does not depend so intensely on constant approval of new type designs, potentially explaining Boeing's willingness to attempt compliance evasion in 737 MAX certification. US Department of Justice, "Boeing Charged with 737 MAX Fraud."
[198] For more on the lack of fundamental understanding of machine learning systems, see Dario Amodei et al., "Concrete Problems in AI Safety," arXiv, 2016, https://doi.org/10.48550/arXiv.1606.06565; Jiaming Ji et al., "AI Alignment: a Comprehensive Survey," arXiv, 2024, https://doi.org/10.48550/arXiv.2310.19852.
[199] Guha et al., "AI Regulation Has Its Own Alignment Problem," 46.
[200] SpaceX's Falcon rockets, for example, regularly launch and land to place satellites in orbit, among other tasks. See Lars Blackmore, "Autonomous Precision Landing of Space Rockets," *The Bridge* 46, no. 4 (2016): 15-20, https://www.researchgate.net/publication/316547389_Autonomous_precision_landing_of_space_rockets.
[201] "Changing the Collective Risk Limits for Launches and Reentries and Clarifying the Risk Limit Used to Establish Hazard Areas for Ships and Aircraft," Federal Aviation Administration, 81 Fed. Reg. 47017 (2016), https://www.federalregister.gov/documents/2016/07/20/2016-17083/changing-the-collective-risk-limits-for-launches-and-reentries-and-clarifying-the-risk-limit-used-to.

The true bind of quantification is the "inapposite mapping between the world of AI and the world of large samples and well-defined risks in which approval regulation operates."[204] The example of pharmaceuticals is instructive. Over a century of regulatory practice, including many examples of harms and near-harms caused by drugs, has accumulated a substantial amount of data upon which to base risk estimates and other factors relevant to approval. For instance, the likelihood of a specific drug causing long-term heart problems may be readily calculable from historical data. Indeed, the FDA employs an Office of Biostatistics, whose main role is to oversee and verify the use of statistical methods throughout the approval process.[205] Frontier AI simply has not generated data in the same way. Although it is true that evaluations can in theory test for any behavior, these can be unreliable[206] and risks can be presented over and above evaluation behavior.[207]

It may be possible, however, to effectively regulate AI without such precise quantification of risks. Some propose, for example, the reduction of risk to a level "as low as reasonably practicable" so long as that risk is generally judged to be below a clearly unacceptable level.[208] Specifications for safety should be designed with the difficulty of quantifying risk in mind. On the other hand, more reliable methods may be developed to make accurate statements about the quantitative risks associated with models. For example, a "sandbox environment" (a protected and secure testing environment for models) might facilitate safe large-scale data collection through behavioral evaluation, approximating years of FDA or FAA experience.[209] In addition, further research may provide insight into causes[210] and solutions[211] for the unreliability and inconsistency of model responses, which further confounds data collection. Because it has yet to be shown that risk quantification, though difficult, is strictly necessary for reliable certification, this challenge is blurry.

---

[202] For example, "Safety Cases" proposes assigning probabilities to each claim in a goal structuring notation (GSN) graph and then aggregating up the graph to the major claim: that the model is ready for deployment. For instance, regulators and an applicant could assign a 99.7% likelihood that the model 'does not display racist behavior in decision-making' on the basis of an evaluation. Clymer et al., "Safety Cases."

[203] The National Institute of Standards and Technology puts it more simply: "AI risks or failures that are not well-defined or adequately understood are difficult to measure quantitatively or qualitatively." Raimondo, *Artificial Intelligence Risk Management Framework*, 5.

[204] Carpenter and Ezell, "An FDA for AI," 10.

[205] "Office of Biostatistics," US Food and Drug Administration, April 29, 2024, https://www.fda.gov/about-fda/cder-offices-and-divisions/office-biostatistics.

[206] For instance, recent research indicates that evaluations of "AI Agents"—models (and scaffolding) that can make decisions and perform tasks in complex environments—are flawed for multiple critical reasons. Sayash Kapoor et al., "AI Agents that Matter," arXiv, 2024, https://doi.org/10.48550/arXiv.2407.01502.

[207] For example, the helpful language models used today as chatbots have been shown to purposefully deceive evaluators and lie about their nefarious behavior in some experiments. Olli Järviniemi and Evan Hubinger, "Uncovering Deceptive Tendencies in Language Models: A Simulated Company AI Assistant," arXiv, 2024, https://doi.org/10.48550/arXiv.2405.01576.

[208] See Koessler, Schuett, and Anderljung, "Risk Thresholds," section 3.

[209] Such a sandbox has been proposed in the EU. See Madiega and Van De Pol, "Artificial Intelligence Act and Regulatory Sandboxes."

[210] See Miao Xiong, "Can LLMs Express Their Uncertainty? An Empirical Evaluation of Confidence Elicitation in LLMs," arXiv, 2024, https://doi.org/10.48550/arXiv.2306.13063; Nino Scherrer et al., "Evaluating the Moral Beliefs Encoded in LLMs," arXiv, 2023, https://doi.org/10.48550/arXiv.2307.14324.

[211] See Potsawee Manakul, Adian Liusie, and Mark J. F. Gales, "SelfCheckGPT: Zero-Resource Black-Box Hallucination Detection for Generative Large Language Models," *Conference on Empirical Methods in Natural Language Processing* (2023): 9004-9017, https://aclanthology.org/2023.emnlp-main.557.pdf; Sebastian Farquhar, "Detecting Hallucinations in Large Language Models Using Semantic Entropy," *Nature* 630 (2024): 625-30, https://doi.org/10.1038/s41586-024-07421-0.

Basis specification

Perhaps the most threatening of all challenges to pre-market approval of AI is the difficulty of specifying a set of line items which, taken together with confidence, beget deployment readiness. The creation of a reliable and general Certification Basis (CB) cannot be taken for granted. For one thing, unique requirements are needed for each possible deployment environment, spanning open source, public query (i.e., API) access, limited deployment, internal deployment, military use, and others. In addition, requirements are likely to change over time, since specific conditions on models are unlikely to be relevant over the sea change the industry undergoes every few years. For example, virtually all benchmarks in frontier AI are saturated (i.e., so solved as to become useless for measurement) within a few years.[212] Yet, the sheer scale and dynamism of any standardized certification requirements can be seen as something of a solvable resource issue. After all, FAA requirements for certification of "Transport Airplanes" alone span thousands of specific, detailed conditions, but these are used in every relevant process with acuity.[213] The deeper issue with basis specification is the reliability of threat model enumeration.

 Threat modeling, as discussed in compliance planning, involves ensuring the safety of a system by listing the avenues through which it could cause harm. The FAA can be seen as doing something akin to threat modeling in determining its type certification requirements. For instance, one threat model is that, in one of the massive number of aircraft flights, the pilots become unresponsive and unable to fly the aircraft.[214] The corresponding certification requirement is that "There must be an emergency means to enable a flight attendant to enter the pilot compartment in the event that the flightcrew becomes incapacitated."[215] Comprehensive threat enumeration is the leading theory for a method of ensuring model safety in something like a CB.[216] However, there are reasons to think doing this is more difficult than in the established cases.

 For one, the risks associated with frontier AI are dramatically poorly understood. This technology is merely a few years old, whereas experimentation—and experience of real large-scale harms—with pharmaceuticals and aircraft are many decades familiar. This leads to significant risk of unknown unknowns: harms that are not foreseen until they occur. Second, regulators may fall prey to the "Lucretius problem": a failure in which standards are set under the false belief that no harms substantially different or more damaging than those experienced in the past can come about.[217] Third, and perhaps most pressing, decomposition into threat models may fundamentally not correctly collate into a robust assurance of deployment readiness. This is most likely to happen because of coordinated (i.e., "normal") failures between threats[218] or because the scope of threats is too large to adequately address in a finite CB.

---

[212] See Simon Ott et al., "Mapping Global Dynamics of Benchmark Creation and Saturation in Artificial Intelligence," *Nature Communications* 13 (2022), https://doi.org/10.1038/s41467-022-34591-0; "Artificial Intelligence Index Report 2023: Chapter 2: Technical Performance," Stanford Human-Centered Artificial Intelligence, 2023, 5, 31-40, 46, https://aiindex.stanford.edu/wp-content/uploads/2023/04/HAI_AI-Index-Report-2023_CHAPTER_2.pdf.
[213] Airworthiness Standards: Transport Category Airplanes, 14 C.F.R., §25.
[214] This is, coincidentally, the premise of the 1980 comedy film *Airplane!*, in which food poisoning incapacitates the flight crew and forces an unsuspecting passenger to make a landing. *Airplane!* Directed by Jim Abrahams, David Zucker, and Jerry Zucker (Hollywood, CA: Paramount Pictures), 1980.
[215] Airworthiness Standards: Transport Category Airplanes, 14 C.F.R., §25.772.
[216] Comprehensive enumeration is suggested, for example in "Safety Cases," as the preferred method of bearing out a claim related to deployment readiness. Clymer et al., "Safety Cases," 41-2.

Basis specification is unlikely to succeed unless significant future work is invested in improving its methods. For this reason, it is a deep challenge.

### Showing compliance

For the most part,[219] the CB is developed prior to an applicant's project process. The specific task of pre-deployment approval regulation is to *show* that the line items in the CB are fulfilled by the relevant model and surrounding assurance measures. This phase is easily the most demanding within the established approval regulation regimes. Clinical trials, for instance, can collectively take as long as a decade.[220] Aircraft type certification tests require thousands of hours of flight time and many more on the ground or with individual parts.[221] In addition, both the FDA and FAA perform site inspections and other methods for verification.[222] In the case of frontier AI, the question is how reliably evaluation, safety, and security measures can be created and implemented. Each of these methods is an input to the overall ability of the applicant to fulfill the CB line items in a way that is verifiable to the FAIA. Safety and security ensures deployment readiness, while evaluation demonstrates it.

Significant work is pouring into each of these inputs. New research into the science of evaluations has sprung up, asking what makes an effective and useful evaluation of a model, particularly those which are increasingly capable and agentic (i.e., able to take complete actions). The National Institute of Standards and Technology's (NIST) 2024 "Assessing Risks and Impacts of AI program," for example, "aims to improve the quality of risk and impact assessments for the field of safe and trustworthy AI."[223] Evaluations are also a focus for frontier AI developers[224] and additional federal agencies.[225] Research on methods for ensuring the safety of frontier systems is even a broader field than evaluation, with significant contributions and research programs from large AI firms,[226] academia,[227] and other organizations.[228] Lastly, a cybersecurity focus has arisen in recent years, to the extent that, for example, OpenAI undergoes "third-party penetration testing" to improve its security.[229] This is in addition to a renewed focus on cybersecurity from a policy and national security perspective.[230]

---

[217] This issue is raised with respect to AI in Carpenter and Ezell, "An FDA for AI," 6.
[218] Such coordinated failures in high-risk industries due to unexpected interactions between seemingly robust subsystems, including in the aviation industry, are discussed at length in Charles Perrow, *Normal Accidents: Living with High-Risk Technologies - Updated Edition* (Princeton University Press, 1999).
[219] With the exception of special considerations, which are discussed at length in the schematic.
[220] "Step 3: Clinical Research," US Food and Drug Administration, January 4, 2018, https://www.fda.gov/patients/drug-development-process/step-3-clinical-research.
[221] See Airbus, "Test and Certification."
[222] See "Type Inspection Report Instructions," Federal Aviation Administration, accessed July 26, 2024, https://www.faa.gov/sites/faa.gov/files/other_visit/aviation_industry/designees_delegations/resources/FAA_Form_8110-31.pdf; "FDA Site Inspections," Northwestern University, accessed July 26, 2024, https://irb.northwestern.edu/compliance-education/fda-site-inspections.html.
[223] Further: "Long-term programmatic outcomes may include guidelines, tools, evaluations methodologies, and measurement methods." Reva Schwartz et al., "The Draft NIST Assessing Risks and Impacts of AI (ARIA) Pilot Evaluation Plan," June 5, 2024, 4, https://ai-challenges.nist.gov/aria/docs/evaluation_plan.pdf. See also Boming Xia et al., "An AI System Evaluation Framework for Advancing AI Safety: Terminology, Taxonomy, Lifecycle Mapping," arXiv, 2024, https://doi.org/10.48550/arXiv.2404.05388; Shavit et al., "Practices for Governing Agentic AI Systems."
[224] See Ethan Perez et al., "Discovering Language Model Behaviors with Model-Written Evaluations," arXiv, 2022, https://doi.org/10.48550/arXiv.2212.09251; Deep Ganguli et al., "Challenges in Evaluating AI Systems," October 4, 2023, https://www.anthropic.com/news/evaluating-ai-systems; Laura Weidinger et al., "Sociotechnical Safety Evaluation of Generative AI Systems," arXiv, 2023, https://doi.org/10.48550/arXiv.2310.11986; Shevlane et al., "Model Evaluation for Extreme Risks."

Nevertheless, these efforts are far from a complete guarantee of safe and verifiable frontier AI. The reliability of these methods remains a salient question.[231] In addition, the current patchwork approach to developing safety and evaluation techniques may prove a challenge for mapping experimental results to specific CB line items. That challenge may place constraints on the sorts of requirements that can be placed on a CB, further increasing the difficulty of basis specification. For these reasons, showing compliance is a deep challenge.

## Resource Challenges

Approval regulation is clearly an involved process. For instance, the $7.2 billion dollar 2024 FDA budget includes "$4.6 billion for medical product safety activities."[232] Given the medical safety responsibilities of the FDA include around 30,000 products and 40,000 production and development facilities,[233] a frontier AI regulator would hardly require a comparable budget. In addition, FDA resources are partially composed of "user fees," which are fees levied upon firms marketing products under FDA jurisdiction.[234] These are generally restricted to specific goals such as reducing regulatory overhead in the approval process. Slightly more than half ($2.4 billion in

---

[225] For example, the first goal in the mission of the US AI Safety Institute is to "[advance] AI safety science through research, including testing, evaluation, validation, and verification of increasingly capable AI models, systems, and agents." This research focus is directly applicable to compliance showing. "The United States Artificial Intelligence Safety Institute: Vision, Mission, and Strategic Goals," US AI Safety Institute, May 21, 2024, 4, https://www.nist.gov/system/files/documents/2024/05/21/AISI-vision-21May2024.pdf. Additionally, the NIST AI Risk Management Framework sets a standard for validation of AI systems, in which "confirmation, through the provision of objective evidence, that the requirements for a specific intended use or application have been fulfilled" is needed. This, too, is precisely the task of showing compliance. Raimondo, "Artificial Intelligence Risk Management Framework," 13.
[226] See "Researching at the Frontier," Anthropic, accessed July 26, 2024, https://www.anthropic.com/research; "Developing Beneficial AGI Safely and Responsibly," OpenAI, accessed July 26, 2024, https://openai.com/safety/.
[227] See Helen Toner and Ashwin Acharya, "Exploring Clusters of Research in Three Areas of AI Safety," 2022, https://cset.georgetown.edu/wp-content/uploads/Exploring-Clusters-of-Research-in-Three-Areas-of-AI-Safety.pdf.
[228] For example, the Center for AI Safety and the Alignment Research Center.
[229] "Security & Privacy," OpenAI, accessed July 26, 2024, https://openai.com/security-and-privacy/.
[230] See Nevo et al., *Securing AI Model Weights*.
[231] See Richard Ngo, Lawrence Chan, and Sören Mindermann, "The Alignment Problem from a Deep Learning Perspective," arXiv, 2024, https://doi.org/10.48550/arXiv.2209.00626; Abeba Birhane et al., "AI Auditing: The Broken Bus on the Road to AI Accountability," arXiv, 2024, https://doi.org/10.48550/arXiv.2401.14462.
[232] "Fiscal Year 2024 Food and Drug Administration Justification of Estimates for Appropriations Committees," US Food and Drug Administration, 2024, 5, https://www.fda.gov/media/166182/download?attachment.
[233] Specifically, as of 2023, the FDA currently approves around 20,000 prescription drugs; 6,500 medical device products; and 1,600 animal drug products. In addition, registered facilities include 10,508 for human drugs; 25,901 for medical devices; and 1,681 for animal drugs. "FDA at a Glance," US Food and Drug Administration, April, 2023, https://www.fda.gov/media/176816/download.
[234] "FDA: User Fees Explained," US Food and Drug Administration, May 22, 2024, https://www.fda.gov/industry/fda-user-fee-programs/fda-user-fees-explained.

2024) of the FDA medical product safety budget derives from user fees.[235] A similar program could be used in frontier AI to service industry and public interests. Care should be taken to ensure that the use of user fees does not carry with it risks of regulatory capture.[236] Regardless, depending on the resources allocated to the FAIA, talent acquisition or an explosion in the number of firms to regulate may present issues for effective regulation.

### Talent acquisition

Public service struggles to attract the best talent. 15% of Americans work in government, but for PhD holders this number is only around 9%.[237] For AI PhDs, this number is lower: about 5%.[238] Possible explanations abound, though a 2020 Center for Security and Emerging Technology report identifies that government jobs "rank lowest [compared to academia and industry] in terms of growth opportunities, colleagues and professional culture, and the ability to pursue research interests—three highly motivating factors among AI PhDs" as principal causes.[239] Security requirements, such as background checks and security clearances, may also deter candidates.[240] The Pentagon has faced a similar technology talent problem for over a decade, as cybersecurity professionals are recruited to large firms and startups offering better pay and benefits. As then-acting Defense Secretary Patrick Shananah said in 2019: "We really get out-recruited."[241]

There are some reasons to think that securing AI talent in the FAIA is critical for effective approval regulation. The FAA's Aircraft Certification Service is an example: it employs over 1,400 "engineers, scientists, inspectors, test pilots, and other experts" to perform and oversee type certification.[242] It is difficult to imagine an agency without significant technical expertise reliably replicating, verifying, and overseeing experiments and arguments about complex AI models in such a fast-moving industry.

There are, however, some reasons to think talent may not be as troubling an issue as it seems. For one thing, there are important ways in which the regulator has a leg up on firms. These include non-disclosable knowledge of processes and experimental results from past certified models, which can be used to inform regulatory decisions. The regulator also learns over time by accruing a private bank of exper-iments, generation and verification strategies, production site certifications, and views on individual firm compliance and reliability.

---

[235] US Food and Drug Administration, "Fiscal Year 2024," 5.
[236] This question is discussed in Lenhart and Myers West, *Lessons from the FDA for AI*, 32-3.
[237] Fiona Hill, "Public Service and the Federal Government," *Brookings*, May 27, 2020, https://www.brookings.edu/articles/public-service-and-the-federal-government/; Jean Opsomer et al., "U.S. Employment Higher in the Private Sector than in the Education Sector for U.S.-Trained Doctoral Scientists and Engineers: Findings from the 2019 Survey of Doctorate Recipients," *National Center for Science and Engineering Statistics*, 2019, table 1, https://ncses.nsf.gov/pubs/nsf21319.
[238] Catherine Aiken, James Dunham, and Remco Zwetsloot, "Career Preferences of AI Talent," June, 2020, 6, https://cset.georgetown.edu/wp-content/uploads/Career-Preferences-of-AI-Talent.pdf.
[239] Aiken, Dunham, and Zwetsloot, "Career Preferences," 13.
[240] Meghan Loomis, Claudio Garcia, and Courtney Csik, "Challenges of Talent Acquisition for Federal Agencies," September 12, 2023, https://www.bakertilly.com/insights/challenges-talent-acquisition-federal-agencies.
[241] Patrick Kelley, "The Pentagon Is Facing a Serious Workforce Problem," July 10, 2019, https://www.govtech.com/workforce/the-pentagon-is-facing-a-serious-workforce-problem.html.
[242] "Aircraft Certification Service (AIR)," Federal Aviation Administration, July 21, 2023, https://www.faa.gov/about/office_org/headquarters_offices/avs/offices/air.

Second, it may be possible to pass off much of the process workflow to applicants, for example by requisitioning automated evaluations from more trusted developers. This type of ecosystem-supporting activity is a potential use for user fees. This would imply that only high-level supervisors, in particular key decision makers on approvals, would need rare technical expertise.

Third, the regulator benefits from the fact that few certification processes are likely to be active at any one time, given the massive resource investments required to pass the compute threshold filter. This allows for a smaller workforce, reducing talent needs. Fourth, technical experts not affiliated with industry firms, such as those in academia, may be brought onto specific projects as technical consultants. Lastly, actions are already being taken, specifically with regards to AI development, to ensure talent comes to and stays in America. The 2023 AI Executive Order, for instance, dictates for "a program to identify and attract top talent in AI" and to "recruit and retain… highly skilled professionals" in AI.[243]

Because the effects of talent constraints may be mitigable, but the status quo is industry domination of talent, this challenge is blurry.

Many involved firms

A further challenge, indeed one that, if granted, exacerbates most other challenges of enforcement and resources, is that the number of firms to be regulated may be "so large as to defy manageability."[244] Clearly, an increase in the number of firms—and especially increases in the number of relevant production sites and the diversity of model designs—would correspond with a decline in the reliability of regulatory scrutiny on each project. Fortunately, the current and projected development landscape involves few firms due to the size of the compute threshold filter.[245] This not only means there are fewer inspections and developers to keep track of, but also that the number of concurrent certification processes at a given time is significantly lower. Few concurrent projects loosens talent requirements. Most importantly, this allows effective scrutiny to be carried out more quickly, reducing regulatory overhead for firms and allowing for swifter production-to-deployment timelines. As it stands, the massive cost to train a single frontier model—potentially $1 billion by 2027[246] and $10 billion by 2032[247]—prohibits the sustenance of an industry with many firms, meaning this is a shallow challenge.

Compute threshold decay and compute scaling

The preceding view that few firms will be able to exist at the true frontier of AI is correct, but does not take into consideration that today's frontier models are tomorrow's lightweights. The compute threshold, though the most effective filter that can be used prior to training a model, is merely a proxy measure for the potential capabilities, and thus risks, presented by a model.[248] Because the efficiency of AI algorithms improves exponentially,[249] models which have the capabilities implied by the $10^{26}$ FLOPs threshold today could be trained using significantly less compute in the future, potentially avoiding the certification process despite having dangerous capabilities.[250] The efficiency of computational resources per dollar also grows exponentially over time, meaning that it will become cheaper to train a model which passes the filter.[251] This implies that a stagnant compute threshold opens the certification process to many more firms over time, increasing its workload and threatening the reliability and expediency of approvals.

---

[243] Exec. Order No. 14110, §10.2.
[244] Carpenter and Ezell, "An FDA for AI," 4.
[245] For analysis, see Paul Scharre, *Future-Proofing Frontier AI Regulation*, Technology and National Security (Washington, DC: CNAS, 2024), https://www.cnas.org/publications/reports/future-proofing-frontier-ai-regulation.
[246] Cottier et al., "The Rising Costs of Training Frontier AI Models," 6.
[247] Scharre, *Future-Proofing*, 2.

These two effects—exponential compute growth and cost decreases—combine to paint a startling picture: despite today costing over $150 million to train a model with capabilities at the compute threshold, by 2028 it will cost only $1 million. Total cost will decrease below $100,000 before 2030.[252] Any regulator in such a situation is pulled in two directions: it seems desirable to *increase* the compute threshold, such that its workload is decreased, but it also seems necessary to *decrease* the compute threshold, such that all potentially dangerous models are covered by the certification process.

Given a single data center-grade AI chip is in the range of $30,000,[253] it seems unreasonable for the FAIA not to increase the threshold; otherwise, it would be regulating any actor with even a few chips at its disposal. In contrast, the certification process schematic as presented in this report is designed with the largest-scale projects of the largest firms in mind. One potential avenue for addressing this issue is a tiered system in which increasing thresholds[254] require increasing scrutiny and FAIA involvement. These could range from merely reporting requirements to the full model certification process. Another solution involves raising the general

---

[248] See Koessler, Schuett, and Anderljung, "Risk Thresholds."

[249] In particular, "the compute required to reach a set performance threshold has halved approximately every 8 months." Anson Ho et al., "Algorithmic Progress in Language Models," arXiv, 2024, 1, https://doi.org/10.48550/arXiv.2403.05812.

[250] Because no current models are known or perceived to have been trained above this threshold, it is not yet clear whether they threaten significant harms, but the general idea that dangerous models may eventually be filtered out by a stagnant threshold still stands.

[251] Computational performance per dollar doubles roughly every 2 years on ML-optimized data center GPUs. Marius Hobbhahn, Lennart Heim, and Gökçe Aydos, "Trends in Machine Learning Hardware," *Epoch AI*, November 9, 2023, https://epochai.org/blog/trends-in-machine-learning-hardware. For more on the description and consequences of the distribution of capabilities due to compute scaling, see Konstantin Pilz, Lennart Heim, and Nicholas Brown, "Increased Compute Efficiency and the Diffusion of AI Capabilities," arXiv, 2024, https://doi.org/10.48550/arXiv.2311.15377.

[252] Scharre, *Future-Proofing*, 33.

[253] Nvidia's flagship AI chip, the H100, "is priced at a massive $30,000." Doug Eadline, "Nvidia H100: Are 550,000 GPUs Enough for this Year?" August 17, 2023, https://www.hpcwire.com/2023/08/17/nvidia-h100-are-550000-gpus-enough-for-this-year/. The firm's upcoming next-generation chip, the B100, will sell at a similar price. Kif Leswing, "Nvidia's Latest AI Chip Will Cost More than $30,000, CEO Says," *CNBC*, March 19, 2024, https://www.cnbc.com/2024/03/19/nvidias-blackwell-ai-chip-will-cost-more-than-30000-ceo-says.html

[254] In a tiered system, thresholds informed by information other than compute (e.g., historical performance data of firms) should be used, as noted in Rishi Bommasani, "Drawing Lines: Tiers for Foundation Models," 2023, https://crfm.stanford.edu/2023/11/18/tiers.html.

threshold. Because no model has yet been trained above the threshold,[255] such models might not realistically pose increased danger compared to current models. In this case, a higher threshold may be warranted. The FAIA should be empowered to raise or lower its threshold over time based on the development landscape and learnings from completed and ongoing certification projects. In addition, the threshold filter need not act as the sole determinant of regulatory intensity. Relatively light touch requirements can be used for less risky models (determined based on, for example, firm history and level of innovation).

Nevertheless, correctly filtering out benign models while capturing dangerous ones remains a significant open question for regulation. This is a deep challenge.

## Competition Challenges

The competition challenges ask whether the FAIA can ensure a competitive market and foster innovation while carrying out approval regulation. The US Department of Justice has already signaled concerns about competitive integrity in frontier AI by launching antitrust inquiries into OpenAI, Microsoft, and Nvidia.[256] In addition, though the US and its major international AI competitor, China, have indicated some willingness to cooperate on reducing risk from frontier AI,[257] national security concerns stemming from foreign development and theft risks remain.[258]

### Undue burdens for small and entering firms

Multiple analyses have noted that approval regulation may "concentrate development amongst large organizations and reduce the ability for smaller or emerging companies to compete."[259] One proposed mechanism by which this could occur is through regulatory capture, in which larger firms successfully lobby for their interests against smaller competitors.[260] However, capture is not necessarily the only culprit in product approval that favors "larger, older producers"; more prolific firms are likely to see shorter approval times and a smaller evidential burden.[261] More straightforwardly, smaller firms may not be able to afford to pursue approval if the process is expensive.

Under the intended application of the model certification process to only the largest projects of the largest firms, the number of small firms is likely to be low (although entrants may be more common).[262] On the whole, the entrant issue can be solved with subsidies and a heightened level of FAIA involvement.[263] Subsidies for small and entering firms may be a promising place to direct funding from user fees. Given the ease with which burdens can be allayed, this is a shallow challenge.

---

[255] Although training compute use is not a publicly reported metric for most large models, best estimates place Gemini Ultra as the largest single model in terms of compute use: on the order of $10^{25}$ FLOPs. Jaime Sevilla and Edu Roldán, "Training Compute of Frontier AI Models Grows by 4-5x per Year," *Epoch AI*, May 28, 2024, https://epochai.org/blog/training-compute-of-frontier-ai-models-grows-by-4-5x-per-year.
[256] David McCabe, "U.S. Clears Way for Antitrust Inquiries of Nvidia, Microsoft, and OpenAI," *New York Times*, June 5, 2024, https://www.nytimes.com/2024/06/05/technology/nvidia-microsoft-openai-antitrust-doj-ftc.html.
[257] See Eva Dou, "U.S.-China Talks on AI Risks Set to Begin in Geneva," *Washington Post*, May 13, 2024, https://www.washingtonpost.com/technology/2024/05/13/us-china-ai-talks/; "Scientists Call For International Cooperation on AI Red Lines," FAR AI, March 18, 2024, https://far.ai/post/2024-03-idais-beijing/.
[258] See Sujai Shivakumar, Charles Wessner, and Thomas Howell, "Balancing the Ledger: Export Controls on U.S. Chip Technology to China," February, 2024, https://www.csis.org/analysis/balancing-ledger-export-controls-us-chip-technology-china; Yingyi Ma, "US Security and Immigration Policies Threaten Its AI Leadership," *Brookings*, April 4, 2024, https://www.brookings.edu/articles/us-security-and-immigration-policies-threaten-its-ai-leadership/.
[259] Guha et al., "AI Regulation Has Its Own Alignment Problem," 41.
[260] See Daniel Carpenter and David A. Moss, *Preventing Regulatory Capture: Special Interest Influence and How to Limit It* (Cambridge, UK: Cambridge University Press, 2014), 1-22.
[261] Carpenter, "Protection Without Capture," 613.

### The impact of regulatory overhead

In part because of export controls on AI chips;[264] in part because of talent issues;[265] and in part because of early adoption, among other factors,[266] US frontier AI is globally preeminent. The military applications and potential economic power of capable frontier AI, in both the near- and medium-term future, means that US national security is threatened if America and its allies lose their lead. Curbing Chinese AI is, for example, the stated goal of tightening semiconductor export controls.[267] The regulatory overhead challenge contends that an approval regulation regime—in general or in the form of this schematic in particular—adds time and costs to development that will unacceptably stifle American AI innovation. This has been, for example, the effect of approval regulation in the case of nuclear energy (though with fewer national security implications).[268]

    The main reason some regulatory overhead is to be expected from any approval regulation regime, including this schematic, is that the government's deployment threshold of safety guarantees is higher than that of firms. Thus, firms are prioritized to increase the amount of safety and security testing they perform on the product before training and deployment.[269] This is the major benefit of approval regulation (alongside the fact that this experimentation then has to be reported to and

---

[262] Frontier AI has seen a number of relevant firms enter with significant investment. For example, Elon Musk's AI company xAI acquired a $24 billion valuation before its first product launch. This makes sense given large investment is required to create a model at the frontier. "Elon Musk's xAI Valued at $24 Bln After Fresh Funding," Reuters, May 27, 2024, https://www.reuters.com/technology/elon-musks-xai-raises-6-bln-series-b-funding-round-2024-05-27/.
[263] On pre-development and pre-deployment approval, a prominent white paper on frontier AI regulation writes: "It is important that such requirements are not overly burdensome for new entrants; the government could provide subsidies and support to limit the compliance costs for smaller organizations." Anderljung et al., "Frontier AI Regulation," 21.
[264] See Jenna Moon, "China Is Falling Behind in Race to Become AI Superpower," *Semafor*, May 8, 2024, https://www.semafor.com/article/05/08/2024/china-ai-development-restrictions-huawei-nvidia.
[265] See Ashwin Acharya and Brian Dunn, "Comparing U.S. and Chinese Contributions to High-Impact AI Research," January, 2022, https://cset.georgetown.edu/wp-content/uploads/Comparing-U.S.-and-Chinese-Contributions-to-High-Impact-AI.pdf; Wouter Maes and Alex Sawaya, "How Businesses Can Close China's AI Talent Gap," May 5, 2023, https://www.mckinsey.com/capabilities/quantumblack/our-insights/how-businesses-can-close-chinas-ai-talent-gap.
[266] For a survey of the question, see Graham Allison and Eric Schmidt, *Is China Beating the U.S. to AI Supremacy*, Avoiding Great Power War Project (Cambridge, MA: Belfer Center for Science and International Affairs, Harvard Kennedy School, 2020), https://www.belfercenter.org/sites/default/files/2020-08/AISupremacy.pdf; Nik Martin, "AI: How Far is China Behind the West," July 24, 2023, https://www.dw.com/en/ai-how-far-is-china-behind-the-west/a-66293806.
[267] See Allen, "Choking Off China's Access."
[268] The extent to which overregulation is entirely responsible for the decline of nuclear energy in the United States is unclear. Some point to the widespread stagnation of electricity generation as another key factor. See Rogé Karma, "Nuclear Energy's Bottom Line," *The Atlantic*, May 26, 2024, https://www.theatlantic.com/ideas/archive/2024/05/nuclear-power-climate-change/678483/.
[269] In other words, "the fact that the regulator likely has a higher bar for converting R&D into product launch than does the firm itself means that firms have incentives to conduct more testing than they otherwise would, and adhere more closely to practices specified by the regulator." Carpenter and Ezell, "An FDA for AI," 3.

verified by regulators), but can also stretch out time-to-market. In an industry as fast paced as frontier AI, where no lead is ever longer than a year, the acceptable level of overhead may be quite small. It should be noted, however, that regulatory scrutiny also acts to prevent national security threats from model theft and capability scaling, meaning overhead reduction at the cost of process reliability may not always be prudent.

A few features of frontier AI and the schematic presented in this report provide a starting point for overhead reduction. First of all, the early history of the FAIA is likely to be characterized by few projects, both annually and concurrently. This allows for regulators to plan the approval process as early as the Ideation phase while devoting a larger portion of resources to each project. This also allows high-level decision makers in the FAIA to have more involvement with projects, expediting approval decisions. Second, as regulators and firms accrue experience with the process, it will become much more efficient. The primary reason for this is the accumulation of evaluations, Advisory Circulars, and safety and security measures, which all reduce the burdens of planning and implementation. Median FDA new drug approval review time, for example, has steadily declined from 32.5 months in 1980 to only 11 months in 2022, despite the number of approved drugs increasing significantly over the same period.[270] Third, the schematic provided in this report takes care to map all approval phases and gates directly onto the standard model development pipeline, which reduces strain from regulation.[271]

Regardless of potential cost- and time-cutting measures, it may be difficult to reduce overhead to an acceptable level without compromising the reliability of scrutiny and, in the end, approval decisions. For this reason, regulatory overhead is a deep challenge.

## Fundamental Challenges

Some are sure to ask whether alternative theories of regulation may facilitate better frontier AI regulation. As discussed in the Introduction, the potential large-scale risks associated with the technology seem to warrant 'hard-touch' regulation on the level of gated approvals. One survey of regulatory theories in AI puts it like this: certification "is often used in scenarios where greater transparency to the public (disclosures) or to the government (registration) is deemed an insufficient safeguard against potentially harmful activity."[272] On this basis, simple disclosures or registrations lack teeth and are thus insufficient for the risks presented by AI. This challenge category generally covers the drawbacks of a few further alternative theories and their potential synergies with approval regulation.

Perhaps the most prominent alternative to approval regulation is a theory based on liability. On this theory, the imposition of liabilities due to negligence can be used to internalize risks from model development into firms (since the largest harms, by default, fall on a national or international public). Yet the possible harms of AI are hidden (i.e., unpredictable) to such an extent that liability is unlikely to sufficiently internalize risk. This is because (1) current frontier AI models are so poorly understood and so thoroughly unregulated that building a sufficient case of "provable human negligence," a requirement for applying liabilities, is extremely difficult and (2) current liability law is not designed to handle harms "so large that paying them out would bankrupt the companies responsible," which must be considered given the national security implications of the technology.[273] In addition, liability does not incentivize firms to develop and use novel safety and security techniques, since liabilities are unlikely to be applied to firms that use industry standard measures.[274] This is a significant drawback compared to approval regulation.[275] Although these challenges make it unrealistic to expect liability to sufficiently protect against the various and large-scale risks posed by frontier AI development, liability could certainly be further legally specified as part of a portfolio designed to, for instance, incentivize firm compliance during model certification.[276]

---

[270] Enrique Seoane-Vazquez, Rosa Rodriguez-Monguio, and John H. Powers III, "Analysis of US Food and Drug Administration New Drug and Biologic Approvals, Regulatory Pathways, and Review Times, 1980-2022," *Scientific Reports* 14, February 9, 2024, https://doi.org/10.1038/s41598-024-53554-7.
[271] There are other reasons to think regulatory burdens are not so costly in the case of frontier AI. For example, geographic concentration may make site inspections easier: OpenAI and Anthropic are both headquartered in San Francisco. Also, most experiments on models can be performed much more quickly than something like an aviation flight test—and certainly faster than a clinical trial—once implemented.
[272] Guha et al., "AI Regulation Has Its Own Alignment Problem," 41. Parentheses in original.

An emerging phenomenon that should also be avoided is the so-called 'patchwork' approach to regulation, in which various fractured limits and requirements are instituted without significant thought given to their cohesion and comprehensiveness.[277] Internet privacy laws have essentially developed this way, which has been awkward given the border-defying nature of internet interactions.[278] In addition to potentially being ineffective for deterring harms, especially those that are difficult to foresee (and thus unlikely to be 'patched'), there is risk of a "regulatory environment that delays access to important products, makes life harder for start-ups,… and undermines responsible development efforts."[279] One positive lesson from the patchwork approach is that regulations should be tightened in the case of observed harms or near-harms. In the approval regulation context, this may mean increased scrutiny during phases of the certification process that incident investigators determine were unable to prevent a failure. Regulators may also benefit from diverse regulatory experimentation given different policies in many states,[280] but this must be viewed as a learning phase, not an effective final governance regime.

---

[273] Gabriel Weil, "Tort Law as a Tool for Mitigating Catastrophic Risk from Artificial Intelligence," SSRN, January 13, 2024, 8, https://dx.doi.org/10.2139/ssrn.4694006.
[274] Gabriel Weil, email message to author, July 22, 2024.
[275] Whether this problem, and others raised with regards to a liability-focused governance regime, are solvable is beyond the scope of this report. However, it is clear that significant changes to existing liability statutes would be required to enact the necessary response. For debate on whether liability could serve to comprehensively regulate AI, compare Matthew van der Merwe, Ketan Ramakrishnan, and Markus Anderljung, "Tort Law and Frontier AI Governance," *Lawfare*, May 24, 2024, https://www.lawfaremedia.org/article/tort-law-and-frontier-ai-governance; Gabriel Weil, "Tort Law Should Be the Centerpiece of AI Governance," *Lawfare*, August 6, 2024, https://www.lawfaremedia.org/article/tort-law-should-be-the-centerpiece-of-ai-governance.
[276] For further discussion of liability law in AI, see Weil, "Tort Law as a Tool"; Tambiama Madiega, "Artificial Intelligence Liability Directive," February, 2023, https://www.europarl.europa.eu/RegData/etudes/BRIE/2023/739342/EPRS_BRI(2023)739342_EN.pdf; John Kingston, "Artificial Intelligence and Legal Liability," arXiv, 2018, https://doi.org/10.48550/arXiv.1802.07782.
[277] One clear way that a patchwork method is coming into view is through distributed state-by-state regulations. See Rachel Wright, "Artificial Intelligence in the States: Emerging Legislation," December 6, 2023, https://www.csg.org/2023/12/06/artificial-intelligence-in-the-states-emerging-legislation/; Digital Democracy - CalMatters, "SB 1047."
[278] See Thorin Klosowski, "The State of Consumer Data Privacy Laws in the US (And Why It Matters)," *Wirecutter* (blog), September 6, 2021, https://www.nytimes.com/wirecutter/blog/state-of-privacy-laws-in-us/.
[279] Kent Walker, "A Patchwork of Rules and Regulations Won't Cut It for AI," *The Hill*, November 5, 2023, https://thehill.com/opinion/technology/4292625-a-patchwork-of-rules-and-regulations-wont-cut-it-for-ai/.

Another proposal involves establishing so-called "regulatory markets," where "governments require the targets of regulation [i.e., applicant firms] to purchase regulatory services from a private regulator."[281] Though such a regime imagines a rich competitive ecosystem improving the reliability of, for instance, basis specification or compliance showing and finding methods, it has deep-seated problems. First, the scope of this report's schematic involves regulation of only a few firms—initiating no more than a few projects per year—hardly enough to support a competitive market of private regulators.[282] Second, regulatory learning is limited, given the private regulators are not incentivized to share information and may cease operation at any point. Third, and perhaps most importantly, the information regarding model specifications (and, in particular, the model itself, required to reproduce experiments) that must be shared with any regulator in order for effective adjudication on safety to be made is likely too sensitive and valuable to share with small private regulatory firms. Nevertheless, if regulatory markets do survive these significant barriers to adoption, there is no particular reason to think that the private regulators could not be empowered with pre-development and pre-deployment approval powers.

Finally, policymakers may be tempted to neglect risks from models per se and instead only consider regulation on downstream applications.[283] This would be a mistake. Unique risks are posed by models alone. The national security risks posed by the training and retention of a highly capable frontier model are huge, as discussed in the Introduction. Theft of a newly-trained model could occur before any downstream applications are considered. This leaves no way to ensure security of valuable and capable models. In addition, certain deployment contexts, such as firm internal deployment, could not legitimately be covered by downstream regulation. Yet, such deployment presents some of the most substantial risks, given the lack of transparency in the industry. Finally, downstream regulators lack sufficient model access to reliably assess safety.[284] Although downstream regulations may bolster safety and consumer protection guarantees, they should not be viewed as a comprehensive solution to the governance of frontier AI.

---

[280] Industry actors may also prefer comprehensive or *ex ante* regulation in many cases, as well: "*Ex ante* regulation… provides consistency and reliability for industry players and regulators alike, in contrast to the whack-a-mole approach that constitutes AI regulation at present." Lenhart and Myers West, *Lessons from the FDA for AI*, 4.
[281] Gillian K. Hadfield and Jack Clark, "Regulatory Markets," 1.
[282] This concern is raised in Gillian K. Hadfield and Jack Clark, "Regulatory Markets," 18-9.
[283] The FDA, for example, has already approved over 800 "AI/ML-enabled medical devices." "Artificial Intelligence and Machine Learning (AI/ML)-Enabled Medical Devices," US Food and Drug Administration, May 13, 2024, https://www.fda.gov/medical-devices/software-medical-device-samd/artificial-intelligence-and-machine-learning-aiml-enabled-medical-devices.
[284] Carpenter and Ezell, "An FDA for AI," 6.

# Recommendations

While giving due consideration to the challenges posed to an agency attempting to implement a schematic of the sort necessary to protect national security, preserve the individual rights of its citizens, and mitigate current and future risks posed by frontier AI, **this report advocates for earnest consideration of an approval regulation regime in this industry**. This section opens with three recommendations that relevant stakeholders should pursue as soon as practicable, given significant work will be required in these areas before effective approval regulation is possible. The next four recommendations need not be fully implemented immediately, but are likewise necessary to enable approval regulation. They should be seriously pursued before approval regulation is brought to bear on frontier models. A final set of three recommendations act as general lessons for any regulator attempting to ensure the safety or security of frontier AI. These recommendations are a product of the analysis in this report and should be heeded by all oversight actors.

These first three recommendations are urgent priorities for any actor hoping to facilitate effective approval regulation in the next few years.

**Improve evaluation techniques**

*Government agencies, frontier AI firms, and academic research should tackle the challenge of effective evaluation and broader testing of AI models.* The US AI Safety Institute,[285] OpenAI,[286] Anthropic,[287] Google DeepMind,[288] and academic researchers[289] already have a clear focus on designing and improving methods for evaluating the capabilities and risks of frontier models. These methods are incredibly important for the reliability of approval regulation, especially during Compliance Showing and Finding. It is unlikely that an effective Project-Specific Certification Plan can be formed without significant advances in the breadth and depth of evaluations and evaluation science.[290] In order to meet this challenge, relevant stakeholders should continue and expand their efforts to build a broad swath of informative experiments on models. In addition, regulators should begin to accumulate a set of proprietary evaluations which, over time, will act as a significant pathway for regulatory learning and leverage for testing the reliability of firm experiments. For more on this recommendation, see Compliance Showing and Finding and Demonstration Challenges.

---

[285] See "Strategic Vision," US AI Safety Institute, May 21, 2024, https://www.nist.gov/aisi/strategic-vision. In addition, the UK AI Safety Institute's first "core function" is to "Develop and conduct evaluations on advanced AI systems." "AI Safety Institute Approach to Evaluations," UK AI Safety Institute, February 9, 2024, https://www.gov.uk/government/publications/ai-safety-institute-approach-to-evaluations/ai-safety-institute-approach-to-evaluations.

[286] See OpenAI, "OpenAI Safety Update: Sharing Our Practices as Part of the AI Seoul Summit," May 21, 2024, https://openai.com/index/openai-safety-update/.

[287] See Sage Lazzaro, "Anthropic Calls for AI Red Teaming to Be Standardized," *Fortune*, June 13, 2024, https://fortune.com/2024/06/13/anthropic-calls-for-ai-red-teaming-to-be-standardized/.

[288] See Laura Weidinger et al., "Holistic Safety and Responsibility Evaluations of Advanced AI Models," arXiv, 2024, https://doi.org/10.48550/arXiv.2404.14068; Shevlane et al., "Model Evaluation for Extreme Risks"; Weidinger et al., "Sociotechnical Safety Evaluation."

[289] See, e.g., Percy Liang et al., "Holistic Evaluation of Language Models," arXiv, 2023, https://doi.org/10.48550/arXiv.2211.09110; John Burden, "Evaluating AI Evaluation: Perils and Prospects," arXiv, 2024, https://doi.org/10.48550/arXiv.2407.09221.

[290] "Evaluation science" refers to studying how to perform effective, reliable, and insightful evaluations.

## Urgent Recommendations

**Improve Evaluation Techniques**

Government agencies, frontier AI firms, and academic research should tackle the challenge of effective evaluation and broader testing of AI models

**Specify Deployment Readiness Conditions**

Policymakers and other relevant stakeholders should enumerate the conditions required for a model to be safely and securely deployed

**Compute Use Tracking and Detection**

Executive action should enforce compute reporting requirements and collaborate with AI chip designers to create and test on-chip hardware for tracking and identification

## Medium-Term Recommendations

**Minimize Regulatory Overhead**

Government agencies and frontier AI firms should research causes of regulatory overhead in approval regulation and methods for reducing them in any application to AI

**Establish Whistleblower Protections**

Through federal action, enhanced whistleblower protections and rewards should be provided to those working in frontier AI firms

**Determine Best Model Filtering Practices**

Government agencies and independent or academic organizations should direct research towards determining the adequacy of and alternatives to a compute use filter

**Establish Information and Personnel Security**

Any regulator should implement reliable information and personnel security to protect dangerous or proprietary information

## General Recommendations for Regulation of Frontier AI

**Regulate Through Development and Deployment**

Any set of regulations which hopes to ensure safe and secure AI should be involved at least from training to post-deployment

**Approval Gating in Any Regime**

Any regulatory approach should consider the use of a regulatory gate to promote transparency, compliance, extra safety and security testing, etc.

**Consider Checkpoint Estimation**

Checkpoint capability estimation should be considered for use in any training oversight regime in order to determine the extent to which developing capabilities are expected and manageable

*A summary of the ten recommendations made in this report. The first layer consists of urgent priorities for the implementation of approval regulation. To succeed, these must be pursued early and earnestly. The second layer gives four priorities for stakeholders that, though less urgent, this report recommends are pursued before approval regulation is used. The final layer has general recommendations that should be heeded by any actor attempting to implement effective, comprehensive regulation of frontier AI.*

### Specify deployment readiness conditions

*Policymakers and other relevant stakeholders should enumerate the conditions required for a model to be safely and securely deployed.* The model certification process cannot be expected to succeed without work beginning early to draft the regulatory codes that will make up the Certification Basis (CB). The current preferred approach, threat modeling, would require significant further work by researchers and policymakers. Otherwise, such a CB would not capture all avenues for risk. Thus, early work should be undertaken to fully expound the line item requirements that might be sufficient, if shown with confidence, to indicate the deployment readiness of a model. Additional work by researchers may tackle the question of whether threat modeling can guarantee safety and security even if given sufficient resources, or whether another approach is needed. For more, see Training and Demonstration Challenges.

**Bolster compute use tracking and detection**

*Executive action should enforce compute reporting requirements and collaborate with AI chip designers to create and test on-chip hardware for tracking and identification.* Regulator insight into the locations and operations of frontier AI-capable data centers is crucial to enforcement of development and deployment restrictions. Current measures dictated by executive order, which require "existence," "location," and "total computing power" of data centers with "computing capacity of $10^{20}$ integer or floating-point operations (FLOPs) per second for training AI," should be strengthened or at least actively enforced.[291] Enforcement may involve site inspections of new large data centers; strengthened measures could include required Know Your Customer practices for cloud computing providers. Further proposals, which would have frontier AI chips include security and workload transparency modules, may enhance the techniques available to regulators.[292] Such a proposal would unlock tools including operation licenses tied to a Training Authorization, usage verification, and operation limitations to exported or smuggled chips.[293] This work would also improve the enforceability of chip and AI export controls. For more, see Enforcement Challenges.

The next four recommendations are less urgent than the above. However, they are significant areas for future research and development. These should be pursued ardently as approval regulation becomes more plausible. Each of these areas of focus should be implemented around the first certification projects and improved over the entire lifespan of the approval regulator.

**Minimize potential regulatory overhead**

*Government agencies and frontier AI firms should research causes of regulatory overhead in approval regulation and methods for reducing them in any application to AI.* Regulatory overhead is a component of approval regulation in which firms and the regulator have aligned interests. Regulators want to reduce overhead (within the bounds of decision reliability) in order to retain the preeminence of the US and its allies on AI; firms want to reduce overhead to bring products to market more cheaply and expeditiously. Though the schematic given in this report takes measures to reduce overhead, this core challenge requires further exploration. NIST or other agencies, in addition to frontier AI firms, should commission work which aims to diagnose and ameliorate regulatory overhead which might result from the implementation of approval regulation for AI. For more, see Competition Challenges.

**Establish whistleblower protections**

*Federal action should provide enhanced whistleblower protections and rewards to those working in frontier AI firms.* Unsanctioned deployment and enforcement issues

---

[291] Exec. Order No. 14110.
[292] See Aarne, Fist, and Withers, *Secure, Governable Chips*; Kulp et al., *Hardware-Enabled Governance Mechanisms*; Shavit, "What Does It Take to Catch a Chinchilla?"; Choi, Shavit, and Duvenaud, "Tools for Verifying."
[293] Aarne, Fist, and Withers, *Secure, Governable Chips*, 10.

pose deep challenges to the implementation of approval regulation. One large step towards the feasibility of enforcement would be to provide enhanced whistleblower protections and rewards to employees of frontier AI firms. Employees fear retaliation, which can be significant in the form of relinquished equity, preventing them from notifying the US government in the case of dangerous activities. Though ordinary protections would cover reporting of illegal activity once a regulatory regime is implemented that specifies illegal actions, the novel risks posed by AI may require additional protections or rewards. Primarily, protections should guarantee that companies "will not enter into or enforce any agreement that prohibits 'disparagement' or criticism of the company for risk-related concerns, nor retaliate for risk-related criticism by hindering any vested economic benefit."[294] For more, see Enforcement Challenges.

### Determine best model filtering practices

*Government agencies and independent or academic organizations should direct research towards determining the adequacy of and alternatives to a compute use filter.* Determining which models should apply to the model certification process is a delicate task. The use of a training FLOPs threshold, proposed in this schematic, is troubled by the scaling of compute and algorithms. The optimal actions of the regulator—specifically, to raise or lower the threshold—should be studied in agency and independent work. In addition, work should aim to recommend the best initial threshold based on risk assessments. The $10^{26}$ metric floated in this report is a tentative recommendation. It is plausible that a higher threshold is sufficient, though more research is needed to make this determination.[295] For more, see Resource Challenges.

### Establish information and personnel security

*Any future regulator should implement reliable information and personnel security to protect dangerous or proprietary information, which may be incredibly valuable but must be shared to enable adequate regulatory scrutiny.* The frontier AI industry is fiercely competitive and the success of firms is largely based on their intellectual property. In addition, any regulator which hopes to effectively ensure model safety and security must have access to certain valuable or hazardous information proprietary to specific firms, such as model compute usage, architecture, and training schedule. A competitive industry cannot be sustained if submission of information to the regulator is not accompanied by reasonable guarantees that such information will be protected from theft or release and that employees of the regulator will not transfer such information to competing firms (e.g., by being hired at one). The FAA, for example, has a similar program based on exemptions to the Freedom of Information Act, including measures for "Safeguarding Classified National Security

---

[294] Jacob Hilton et al., "A Right to Warn About Advanced Artificial Intelligence," June 4, 2024, https://right-towarn.ai/.
[295] It is unlikely that, as of now, a lower threshold is needed. Though, see discussion in Resource Challenges as to why this may change over time.

Information" and "[Protecting] Sensitive Unclassified Information."[296] Any regulator in the frontier AI industry should follow the FAA's lead on information security. It should also bolster its personnel security, perhaps through internally-facing cybersecurity measures (e.g., limitations to the number of employees with access to model weights) and 'cooling off' periods between quitting at a government regulator and being hired at a frontier firm. For more, see Enforcement Challenges.

The following three recommendations act as general lessons gleaned from the analysis in this report and can apply to any regulatory regime enacted for frontier AI, whether corporate or government, domestic or international, approval-based or otherwise.

### Regulate throughout development and deployment

*Any set of regulations which hopes to ensure safe and secure AI should be involved at least from training to post-deployment.* The possibility of theft of a frontier model, by state or non-state actors, begins as training starts. Thus, it is not sufficient for American regulation around frontier AI to apply only to the deployment and use of models. Instead, requirements must apply also to the development process, starting at the latest during training. On the other hand, regulations cannot solely apply to the development process. Without deployment and post-deployment regulations, firms can easily avoid scrutiny and provide access to valuable but dangerous models. For instance, a firm could revert the safety training performed on a model to comply with development regulations (which would likely reduce performance) and deploy this harmful version instead. To prevent each of these failures, a comprehensive regulatory strategy on frontier AI should apply from training to post-deployment. For more on protections throughout the model lifecycle, see Before Training, Training, and Post-Deployment.

### Consider approval gating in any regime

*Any regulatory approach should consider the use of a regulatory gate to promote transparency, compliance, extra safety and security testing, etc.* This report's analysis makes clear that the use of a gate tied to regulatory approval on the basis of product scrutiny shifts firm incentives partially towards those of the national interest. It may not be necessary to include many gates and a host of accompanying requirements to receive some of the aforementioned benefits. As such, proposals occupying any place on the regulatory intensity spectrum should consider a gate. For example, a 'soft-touch' regulatory approach may consider a weak approval gate adjudicated solely on the basis of submitted firm records, taken to be true (i.e., without significant verification efforts). A firm in this case may still be incentivized to increase its testing program, since this may expedite or ensure approval. Future work may wish to scrutinize the incentives caused by gates of varying intensity. For more on gating,

---

[296] Federal Aviation Administration, *Freedom of Information Act Program*, FAA Order 1270.1A (Washington, DC: Federal Aviation Administration, 2015), https://www.faa.gov/documentLibrary/media/Order/FAA_Order_1270.1A.pdf.

see The Model Certification Process.

### Consider checkpoint capability estimation

*Checkpoint capability estimation should be considered for use in any training oversight regime in order to determine the extent to which developing capabilities are expected and manageable.* This schematic proposes the use of checkpoint capability estimation, an original oversight mechanism in which firms are asked before training to estimate the capabilities of their models at various stages and these estimates are checked against real evaluated capabilities during training. This approach provides insight into the novelty of developing capabilities and allows for swift course corrections in safety and security responses. These estimates can be imposed at various levels of intensity, for example from fully firm-administered to fully regulator-administered, and thus applied to various regulatory proposals. Further research should establish the reliability and applicability of this method. For more on checkpoint estimation, see Before Training and Training.

# Conclusion

As the capabilities of frontier AI models continue to progress at a rapid clip, and unprecedented investment into single models continues to grow,[297] it is increasingly clear that thoughtful, comprehensive regulation of this critical industry is needed. The schematic of approval regulation for frontier AI outlined in this report gives one option for the hard-touch governance that the emerging high-risk industry demands.

It begins with consultation between the regulator and applicant firm. If the project is on the frontier, each then prepares to assure a safe and secure training process. Once such a plan is approved, training of the model begins. During training, the regulator develops a set of requirements which together constitute deployment readiness, while the applicant plans experiments to demonstrate compliance with these requirements. After training, each of these is enacted, leading to a final certification decision after compliance data is generated, collated into arguments, and verified. If a positive approval decision is made, the model can be deployed in a specified deployment environment, where it is monitored for incidents and compliance over its lifecycle.

Like any nascent proposal, this schematic faces a number of challenges to its feasibility and effectiveness. Most prominent among these are the possibility of firm noncompliance through unsanctioned model deployment, the inability of the regulator to specify attainable and comprehensive conditions for deployment readiness, the difficulty of demonstrating compliance with a regulatory condition, the question of setting and adjusting a compute threshold filter, and the need to minimize regulatory scrutiny without compromising assurances of safety and security. Nevertheless, the recommendations given in this report aim to tackle these challenges. None seem insurmountable or to pose a mortal threat to approval regulation of AI.

The continued security of the United States demands action to secure AI development from theft, confidently establish model safety, and sustain a competitive and peerless domestic frontier AI industry. An approval regulation scheme, such as the one put forth in this report, can make these guarantees.

---

[297] See, e.g., Jonathan Vanian, "Mark Zuckerberg Indicates Meta is Spending Billions of Dollars on Nvidia AI Chips," *CNBC*, January 18, 2024, https://www.cnbc.com/2024/01/18/mark-zuckerberg-indicates-meta-is-spending-billions-on-nvidia-ai-chips.html; Ashley Stewart, "Microsoft Has a Target to Amass 1.8 Million AI Chips by the End of the Year, Internal Document Shows," *Business Insider*, April 17, 2024, https://www.businessinsider.com/microsoft-gpu-targets-1-8-million-ai-chips-this-year-2024-4.

## Acknowledgements

For helpful comments and discussions, the author thanks Zachary Rudolph, Sanyu Rajakumar, Sophia Lloyd George, Gabriel Weil, Melody Gui, Neel Guha, Aryan Shrivastava, Daniel Carpenter, Helena Tran, Amina Anowara, Austin Coffey, Carson Ezell, and especially Keller Scholl. For many lessons, the author thanks his parents, Erin and Doug. Any remaining views and errors are the author's alone. This project was financially supported by the University of Chicago Existential Risk Laboratory.

## Author

Cole Salvador is an undergraduate at Harvard College. He can be reached by email: colesalvador@college.harvard.edu.